\documentclass[11pt, reprint,aps,prd,twocolumn,showpacs,superscriptaddress]{revtex4-2}

\usepackage[colorlinks=true,linkcolor=blue,citecolor=blue,urlcolor=magenta]{hyperref}

\usepackage{amsfonts,mathrsfs,amsmath,amsthm,amssymb}
\usepackage{epsfig}
\usepackage{bm}
\usepackage{array,multirow}
\usepackage{booktabs}
\usepackage[table]{xcolor}
\usepackage{graphicx}
\usepackage{upgreek}

\newcolumntype{A}{>{\columncolor{gray!25}}c}
\newcolumntype{B}{>{\columncolor{gray!25}}l}

\begin{document}  
\title {\bf 
Quantum simulation of strong charge-parity violation and Peccei-Quinn mechanism}

\author{Le Bin Ho} 
\thanks{Electronic address: binho@fris.tohoku.ac.jp}
\affiliation{Frontier Research Institute 
for Interdisciplinary Sciences, 
Tohoku University, Sendai 980-8578, Japan}
\affiliation{Department of Applied Physics, 
Graduate School of Engineering, 
Tohoku University, 
Sendai 980-8579, Japan}

\date{\today}

\begin{abstract}
Quantum Chromodynamics (QCD) admits a topological $\bar{\theta}$ term that violates charge-parity ($CP$) symmetry, yet experiments indicate that $\bar{\theta}$ is extremely small. To investigate this problem in a controlled setting, we derive a Hamiltonian formulation of QCD through a $(1+1)$-dimensional Schwinger-model analogue. Fermionic and gauge degrees of freedom are encoded into qubits using Jordan-Wigner and quantum-link mappings, yielding a compact Pauli Hamiltonian that preserves the essential topological vacuum structure. Ground states are prepared using a feedback-based quantum optimization protocol, providing access to the vacuum energy on few-qubit simulators. We observe vacuum minima at $\bar{\theta}=0$ and $2\pi$, consistent with the continuum QCD expectations within the accessible regime. Upon coupling to a dynamical axion field, the system relaxes to $\theta_{\rm eff}=0$, realizing the Peccei-Quinn mechanism within a minimal quantum simulation. These results demonstrate how quantum simulation can probe $CP$ violation and its dynamical resolution in gauge theories.
\end{abstract}
%
%
%\pacs{03.65.Ta, 03.65.Aa, 02.50.-r, 03.67.Ac }
\maketitle

\section{Introduction} 
%\inlinesection{Introduction}
Quantum chromodynamics (QCD) is an SU(3) non-Abelian gauge theory describing the interactions of quarks and gluons \cite{https://doi.org/10.1002/andp.200051211-1210,Gross2023,doi:10.1142/S0217732324501943}.
In addition to the standard kinetic and interaction terms~\cite{doi:10.1142/S0217732324501943}, the QCD includes a topological term arising from gauge-field configurations with nontrivial winding number. This term violates the charge-parity \((CP)\) symmetry when its coefficient \(\theta\) is nonzero~\cite{PhysRevLett.37.8,PhysRevD.14.3432} and governs key aspects of the nonperturbative vacuum structure~\cite{RevModPhys.82.557}. Including this contribution, the QCD Lagrangian can be written as
\begin{align}\label{eq:LQCD}
\mathcal{L}_{\rm QCD}
=
-\frac{1}{4} G_{\mu\nu}^a G^{a,\mu\nu}
+\sum_f \bar{\psi}_f\left(i\gamma^\mu D_\mu - m_f\right)\psi_f \nonumber\\
+\bar{\theta}\,\frac{g^2}{32\pi^2}\, G_{\mu\nu}^a \tilde{G}^{a,\mu\nu}.
\end{align}
Here
\(G_{\mu\nu}^a=\partial_\mu A_\nu^a
-
\partial_\nu A_\mu^a
+
g f^{abc} A_\mu^b A_\nu^c
\)
denote the gluon field-strength tensors, and
\(
\tilde{G}^{a,\mu\nu}
=
\frac{1}{2}\epsilon^{\mu\nu\rho\sigma} G_{\rho\sigma}^a
\)
are their duals, where \(A_\mu^a\) are the gluon gauge fields and \(f^{abc}\) are the structure constants of the SU(3) Lie algebra.  
The quark field of flavor \(f\) is denoted by \(\psi_f\) and carries mass \(m_f\). It interacts with the gluon field via the gauge-covariant derivative
\(
D_\mu
=
\partial_\mu
-
i g\, T^a A_\mu^a ,
\)
where \(g\) is the coupling constant and \(T^a\) are the generators of the SU(3) color group, satisfying the commutation relations
\(
[T^a, T^b] = i f^{abc} T^c .
\)

Because the QCD gauge fields admit topologically nontrivial configurations, the most general gauge-invariant action allows a term proportional to \(\theta G_{\mu\nu}^a \tilde{G}^{a,\mu\nu}\), where \(\theta\) is the QCD vacuum angle and controls explicit \(CP\) violation.
When the quark mass matrix \(M\) carries complex phases, an anomalous axial \(U(1)_A\) rotation shifts these phases into the topological term.
As a result, the physically observable \(CP\)-violating parameter is
\(
\bar{\theta}=\theta-\arg\!\det(M).
\)
Since \(G_{\mu\nu}^a\tilde{G}^{a,\mu\nu}\) is odd under the \(CP\) transformation, the QCD is \(CP\)-symmetric only when \(\bar{\theta}=0\), any nonzero value of \(\bar{\theta}\) therefore leads to an explicit \(CP\) violation.

In QCD, the $CP$ violation induces observable low-energy hadronic effects, such as a permanent electric dipole moment (EDM) of the neutron~\cite{CREWTHER1979123,POSPELOV2005119}. Precision experiments, however, have found no such signal. The current bound of the neutron EDM is
$|d_n|<1.8\times10^{-26}\,e\cdot\mathrm{cm}$ $(90\%~\mathrm{C.L.})$~\cite{PhysRevLett.97.131801,PhysRevLett.124.081803},
implies an exceptionally small value $|\bar{\theta}|\lesssim10^{-10}$. This strong suppression cannot be explained within the Standard Model, leading to the strong $CP$ problem: although the QCD permits an arbitrary $\bar{\theta}$, nature selects a value extremely close to zero.

A natural resolution of the strong $CP$ problem is provided by the Peccei-Quinn (PQ) mechanism~\cite{PhysRevLett.38.1440,PhysRevD.16.1791,Peccei2008,RevModPhys.82.557}, which promotes $\bar{\theta}$ to a dynamical variable associated with a spontaneously broken global $U(1)_{\rm PQ}$ symmetry. The resulting pseudo-Nambu-Goldstone boson,  or the axion, couples anomalously to gluons,
\begin{align}
\mathcal{L}_{a} = \frac{g^{2}}{32\pi^{2}}\,\frac{a}{f_a}\, G_{\mu\nu}^a \tilde{G}^{a,\mu\nu},
\end{align}
where $f_a$ is the axion decay constant. This interaction shifts $\bar\theta$ to an effective angle 
\(
\theta_{\rm eff}=\bar{\theta}+a/f_a.
\)
Nonperturbative QCD dynamics generate an effective potential $V(\theta_{\rm eff})$ whose minimum lies at $\theta_{\rm eff}=0$, independent of microscopic details. Consequently, the axion acquires a vacuum expectation value $\langle a\rangle=-\bar{\theta} f_a$, dynamically canceling the $CP$-violating phase. After the shift $a=a_{\rm phys}+\langle a\rangle$, the QCD vacuum becomes $CP$ symmetric, while the physical axion $a_{\rm phys}$ remains as a light excitation about the minimum. In this way, the PQ mechanism restores the $CP$ symmetry without fine-tuning, while the axion also provides a well-motivated dark-matter candidate connecting particle physics with cosmological observations~\cite{doi:10.1126/sciadv.abj3618}.

Progress on the strong-$CP$ problem is fundamentally limited by the intrinsically nonperturbative nature of the QCD vacuum. Conventional lattice approaches encounter severe challenges such as  such as the sign problem~\cite{NAGATA2022103991} and topological freezing~\cite{ALLES1996107,Bonanno2024}, while analytical treatments necessarily rely on uncontrolled approximations. These limitations motivate alternative approaches. Quantum simulation~\cite{PRXQuantum.4.027001,PRXQuantum.5.037001,PRXQuantum.3.040316} provides a promising route by realizing a controllable quantum system whose Hilbert space directly encodes the gauge theory, allowing direct probing of vacuum responses to a $CP$-violating background. In this context, the $(1+1)$-dimensional Schwinger model offers an ideal testbed: it retains essential features such as confinement and topological vacuum sectors, yet remains sufficiently compact to be mapped onto a small number of qubits and implemented on near-term quantum hardware. Moreover, this model has been extensively explored in cold-atom quantum simulators \cite{PRXQuantum.3.040316,Zhang2025}, highlighting its experimental accessibility and positioning quantum simulation as a viable path forward for studying strong-$CP$ physics.

In this work, we use the $(1+1)$D Schwinger model to investigate the microscopic origin of strong $CP$ violation and its axionic resolution. We construct a qubit-encoded lattice gauge theory with a $\bar{\theta}$-dependent sector, extend it to include a dynamical axion field, and prepare its vacuum using a feedback-driven quantum optimization algorithm (FALQON)~\cite{PhysRevLett.129.250502,nguyen2025imaginary}. By evaluating the vacuum energy as a function of $\bar{\theta}$, we first characterize the $\bar{\theta}$-dependent vacuum structure in the absence of axions, including deviations induced by reduced dimensionality and finite-size effects. Upon coupling to a dynamical axion, the vacuum energy relaxes to a minimum at $\theta_{\rm eff}=0$ and becomes independent of $\bar{\theta}$. These results demonstrate that the PQ mechanism can be realized in controlled few-qubit quantum simulations, providing a direct route for probing topological gauge dynamics on quantum hardware.

\section{Results}
%\inlinesection{Gauge-field Lagrangian to Hamiltonian}
\subsection{Gauge-field Lagrangian to Hamiltonian}
To express the QCD Lagrangian in a form suitable for canonical quantization, we derive the gluon field-strength tensor \( G_{\mu\nu}^a \) in terms of its temporal and spatial components. Analogous to electromagnetism, we define the  chromoelectric  and  chromomagnetic  fields as
\begin{align}
    E_i^a = G_{0i}^a, \quad B_i^a = -\frac{1}{2}\epsilon_{ijk}G_{jk}^a,
\end{align}
where \(i,j,k\) denote spatial indices, and \(a = 1, \dots, 8\) labels the color components of the SU(3) gauge group. 

Substituting these definitions into the gauge-field term in the QCD Lagrangian \eqref{eq:LQCD}, i.e., 
\(
    \mathcal{L}_{\text{gauge}} = -\frac{1}{4}G_{\mu\nu}^a G^{a,\mu\nu} + \bar\theta\frac{g^2}{32\pi^2}G_{\mu\nu}^a\tilde{G}^{a,\mu\nu},
\)
and using the identity \( G_{\mu\nu}^a\tilde{G}^{a,\mu\nu} = -4\mathbf{E}^a \cdot \mathbf{B}^a \), we obtain %the Lagrangian density expressed in terms of the physical fields
\begin{align}\label{eq:lagH}
    \mathcal{L}_{\text{gauge}} = \frac{1}{2}\mathbf{E}^a \cdot \mathbf{E}^a - \frac{1}{2}\mathbf{B}^a \cdot \mathbf{B}^a + \bar\theta\frac{g^2}{8\pi^2}\mathbf{E}^a \cdot \mathbf{B}^a.
\end{align}
The first two terms represent the kinetic and potential energy densities of the gluon fields, respectively, while the last term introduces the coupling between the chromoelectric and chromomagnetic fields. 

Under the \(C\) and \(P\) transformations, these fields transform as,
\(
%\begin{align}
   \mathbf{E}^a \xrightarrow{P} -\mathbf{E}^a, \ \mathbf{B}^a \xrightarrow{P} \mathbf{B}^a, \
\mathbf{E}^a \xrightarrow{C} -\mathbf{E}^a, \ \mathbf{B}^a \xrightarrow{C} -\mathbf{B}^a. 
%\end{align}
\)
Therefore, the scalar product \(\mathbf{E}^a \cdot \mathbf{B}^a\) is odd under both $P$ and $CP$:
\(
%\begin{align}
    \mathbf{E}^a \cdot \mathbf{B}^a \xrightarrow{CP} -\mathbf{E}^a \cdot \mathbf{B}^a.
%\end{align}
\)
Consequently, the term proportional to \(\bar\theta\mathbf{E}^a \cdot \mathbf{B}^a\) 
explicitly breaks the $CP$ symmetry, and induces observable effects such as a permanent EDM in color-neutral hadrons~\cite{CREWTHER1979123,POSPELOV2005119}.

The canonical momentum conjugate to the spatial gauge field \(A_i^a\) is given by
\begin{align}
    \Pi_i^a = \frac{\partial \mathcal{L}_{\text{gauge}}}{\partial(\partial_0 A_i^a)} = E_i^a + \bar\theta\frac{g^2}{8\pi^2}B_i^a,
\end{align}
showing that the \(\bar\theta\)-term shifts the momentum by an amount proportional to the chromomagnetic field. This term couples the electric and magnetic sectors, thus \(\Pi_i^a\) no longer equals the electric field, and thereby modifying the canonical structure of the theory.

The Hamiltonian is then constructed via the Legendre transformation \cite{10.1119/1.3119512}
\begin{align}\label{eq:Hg}
\notag    \mathcal{H}_{\text{gauge}} &= \Pi_i^a\partial_0 A_i^a - \mathcal{L}_{\text{gauge}}\\
&= \frac{1}{2}\Big(\mathbf{E}^a \cdot \mathbf{E}^a + \mathbf{B}^a \cdot \mathbf{B}^a\Big).
\end{align}
%

%\subsection{Simplified (1+1)D U(1) Schwinger model}
\subsection{Simplified (1+1)-dimensional U(1) Schwinger model}
We consider the $(1+1)$D Schwinger model, consisting of one spatial and one temporal dimension~\cite{PhysRevD.52.6435,PhysRev.128.2425}. Despite its Abelian gauge structure, this model reproduces several key nonperturbative features of QCD, including charge confinement, a dynamically generated mass gap, and spontaneous chiral symmetry breaking in the massless limit~\cite{PhysRev.128.2425,Luo2007,PhysRevD.104.126029,PhysRevD.101.054507,PRXQuantum.3.040316}. It further exhibits a nontrivial topological vacuum structure analogous to the $\bar{\theta}$ vacua of QCD, enabling the studies of $CP$-violating effects induced by a background $\bar{\theta}$ term in $(1+1)$D~\cite{PhysRevD.101.054507,PhysRevD.46.5598}. Crucially, the Schwinger model admits a well-defined lattice discretization and a finite-dimensional gauge-link formulation~\cite{nt76-ttmj}, making it well suited for implementation on quantum computing platforms~\cite{PRXQuantum.2.017003}.

In $(1+1)$D, the gauge field has a single spatial component and the magnetic field vanishes identically. Gauge-field dynamics are therefore governed solely by the electric field $E(x)$, which is canonically conjugate to the spatial gauge potential $A_1(x)$ and satisfies
\(
[A_1(x),E(y)] = i\,\delta(x-y),
\)
analogous to the position-momentum commutation relation in quantum mechanics. The continuum Lagrangian for a single fermion flavor coupled to this Abelian gauge field, including a topological $\bar{\theta}$ term, is
\begin{align}
\mathcal{L}
= \bar{\psi}\!\left(i\gamma^\mu D_\mu - m\right)\!\psi
+\frac{1}{2}E^2
+\frac{\bar{\theta}}{2\pi}E,
\end{align}
with the gauge-covariant derivative defined as $D_\mu=\partial_\mu-i g A_\mu$.
Performing the Legendre transformation yields
\begin{align}
H = \int dx \Big[
\psi^\dagger(x)\big(-i\gamma^0\gamma^1 D_1 + m\gamma^0\big)\psi(x) \nonumber \\
+ \frac{1}{2}\Big(E(x)-\frac{\bar{\theta}}{2\pi}\Big)^2
\Big],
\end{align}
where the shift $E(x)\to E(x)-\bar{\theta}/(2\pi)$ encodes the $CP$-odd topological term. This form plays the same role as the $\mathbf{E}\cdot\mathbf{B}$,but is simpler in one spatial dimension.

For quantum simulation, we employ the Kogut-Susskind lattice discretization~\cite{PhysRevD.11.395,PhysRevD.91.054506},
with gauge links $U_{x,x+1}=e^{iA_{x,x+1}}$ and electric fields $E_{x,x+1}$ on links.
The lattice Hamiltonian reads
\begin{align}
H = \frac{m}{2}\sum_x (-1)^x Z_x &
+ \dfrac{w}{2} \sum_x  \big(\psi_x^\dagger  U_{x,x+1}\psi_{x+1} + \mathrm{h.c.}\big) \nonumber\\
& + \frac{g^2}{2}\sum_x \Big(E_{x,x+1}-\frac{\bar{\theta}}{2\pi}\Big)^2,
\end{align}
where $Z_x=1-2\psi_x^\dagger\psi_x$ is the local Pauli-$Z$ operator (equivalently $Z_x=1-2n_x$ with $n_x=\psi_x^\dagger\psi_x$).
The factor $(-1)^x$ arises from the staggered-fermion formulation, which suppresses fermion doubling.

Gauge invariance is enforced by the Gauss law at each lattice site. We impose this constraint by adding a penalty term
$\lambda\sum_x G_x^2$, with the local generator
\begin{align}
G_x = E_{x-1,x} - E_{x,x+1} - \psi_x^\dagger\psi_x + \eta_x,
\end{align}
where $\eta_x=\frac{1+(-1)^x}{2}$ denotes a background charge. 
Physical states satisfy $G_x|\mathrm{phys}\rangle=0$.

\subsection{Two sites lattice case}
Restricting to the smallest nontrivial system, a two-site lattice connected by a single gauge link, yields the minimal Hamiltonian
\begin{align}\label{eq:Ham_2site}
H = \frac{m}{2} & \big(Z_0 - Z_1\big)
+ \dfrac{w}{2}\big(\psi_0^\dagger U_{01}\psi_1 + \mathrm{h.c.}\big) \nonumber\\
&\quad + \frac{g^2}{2}\left(E_{01}-\frac{\bar{\theta}}{2\pi}\right)^2
+ \lambda \left(G_0^2 + G_1^2\right).
\end{align}
The first term describes the staggered mass energy, the second implements gauge-invariant fermion hopping mediated by the link operator $U_{01}$, and the third gives the electric-field energy with the $CP$-violating $\bar{\theta}$ shift. The final term enforces the Gauss law, confining the dynamics to the gauge-invariant subspace.

This minimal Schwinger model retains the essential topological and $CP$-violating structure of the QCD Hamiltonian. The $\bar{\theta}$-dependent electric field acts as a $CP$-odd background, allowing studies of parity violation, vacuum degeneracy, and $\bar{\theta}$-vacuum transitions in a low-dimensional analogue of strong interactions.

To implement the model on quantum hardware, both the fermionic and gauge degrees of freedom are encoded in qubits. The fermionic modes are mapped to qubits using the Jordan-Wigner (JW) transformation~\cite{Jordan1928}, which preserves the correct anticommutation relations through nonlocal parity strings. The staggered fermion fields $\psi_x$ and $\psi_x^\dagger$ satisfy the canonical anticommutation relations
\begin{align}
\{\psi_x,\psi_y^\dagger\}=\delta_{xy}, \qquad
\{\psi_x,\psi_y\}=0 .
\end{align}

In the JW representation they are expressed as
\begin{align}
\psi_x = \Big(\prod_{y<x} Z_y\Big) S_x^-,
\qquad
\psi_x^\dagger = \Big(\prod_{y<x} Z_y\Big) S_x^+ ,
\end{align}
where $S_x^\pm = (X_x \mp iY_x)/2$ are the qubit ladder operators ($S_x^- = |0\rangle\langle1|$, $S_x^+ = |1\rangle\langle0|$).
The string operator $\prod_{y<x} Z_y$ enforces the fermionic parity required to reproduce the correct anticommutation relations. Here $X_x$, $Y_x$, and $Z_x$ denote the Pauli operators acting on the qubit representing the fermionic mode at site $x$.

The gauge field on each link is encoded using the quantum-link representation. For a single-qubit link $\ell$, the electric-field and link operators are represented as
\begin{align}
E_\ell = \frac{I - Z_\ell}{2}, \qquad
U_\ell = \frac{X_\ell + iY_\ell}{2},
\end{align}
where $X_\ell$, $Y_\ell$, and $Z_\ell$ are Pauli operators acting on the qubit associated with the gauge link. A detailed derivation of this mapping is provided in App.~\ref{appA}, while the extension to multi-qubit gauge links is discussed in App.~\ref{appD}.

Substituting these mappings into the Hamiltonian~\eqref{eq:Ham_2site} yields an explicit Pauli-operator representation. The mass term becomes
\begin{align}
H_m = \frac{m}{2}\big(Z_0 - Z_1\big).
\end{align}

The electric-field energy, including the $CP$-violating $\bar{\theta}$ shift, is
\begin{align}
H_{\bar{\theta}}
&= \frac{g^2}{2}\Big(E_\ell-\frac{\bar{\theta}}{2\pi}\Big)^2 \nonumber\\
&= \frac{g^2}{2}\!\left[\!\left(\frac{\bar{\theta}}{2\pi}\right)^2
- \frac{\bar{\theta}}{2\pi} + \frac{1}{2}\!\right] I_\ell
+ \frac{g^2}{2}\!\left(\frac{\bar{\theta}}{2\pi}-\frac{1}{2}\right) Z_\ell,
\end{align}
where we used $E_\ell = E_{01}$ is the specific link connecting site 0 and site 1.
The identity term produces a uniform energy shift and does not affect the dynamics, while the $Z_\ell$ term changes sign at $\bar{\theta}=\pi$, ensuring the correct $2\pi$ periodicity of the spectrum.

The gauge-invariant fermion hopping term takes the form
\begin{align}
H_{\rm hop}
&= \frac{w}{2}\big(\psi_0^\dagger U_{01}\psi_1 + \psi_1^\dagger U_{01}^\dagger \psi_0\big) \nonumber\\
&= \frac{w}{8}\Big(
X_0 X_1 X_\ell
+ Y_0 Y_1 X_\ell
- X_0 Y_1 Y_\ell
+ Y_0 X_1 Y_\ell
\Big),
\end{align}
which consists of three-body Pauli interactions coupling the two fermionic qubits to the gauge-link qubit. These terms describe correlated hopping processes in which a fermion moves between sites while simultaneously changing the electric flux on the link, thereby preserving local gauge invariance.

The Gauss law is enforced through an energy-penalty term that confines the dynamics to the gauge-invariant subspace 
\begin{align}
H_G = \lambda \big(G_0^2 +  G_1^2\big),
\end{align}
where we choose larger $\lambda \gg \max(m,w,g^2)$ to reduce the gauge-violating excitations.

Combining all contributions, the effective Hamiltonian of the two-site Schwinger model is
\begin{align}\label{eq:Htot}
H_{\rm total} = H_m + H_{\rm hop} + H_{\bar{\theta}} + H_G .
\end{align}

\subsection{The Peccei-Quinn mechanism}
To dynamically relax the $CP$-violating angle, we promote $\bar{\theta}$ to a dynamical variable by introducing an axion field $a(x)$ with decay constant $f_a$.
The axion enters QCD through the shifted effective angle
$\theta_{\rm eff}(x)\equiv \bar{\theta}+a(x)/f_a$.
At the Lagrangian level this corresponds to the replacement
\begin{align}
\bar{\theta}\,\frac{g^2}{32\pi^2}G_{\mu\nu}^a\tilde G^{a,\mu\nu}
\;\longrightarrow\;
\left(\bar{\theta}+\frac{a}{f_a}\right)\frac{g^2}{32\pi^2}G_{\mu\nu}^a\tilde G^{a,\mu\nu},
\end{align}
together with the axion kinetic term $\frac12\,\partial_\mu a\,\partial^\mu a$.
Below the confinement scale, nonperturbative QCD generates an effective potential governed by the topological susceptibility $\chi$,
\begin{align}
V(\theta_{\rm eff}) \simeq \chi\,[1-\cos(\theta_{\rm eff})],
\end{align}
so that $m_a^2=\chi/f_a^2$ in the small-angle limit.

\begin{figure}[b]
    \centering
    \includegraphics[width=\linewidth]{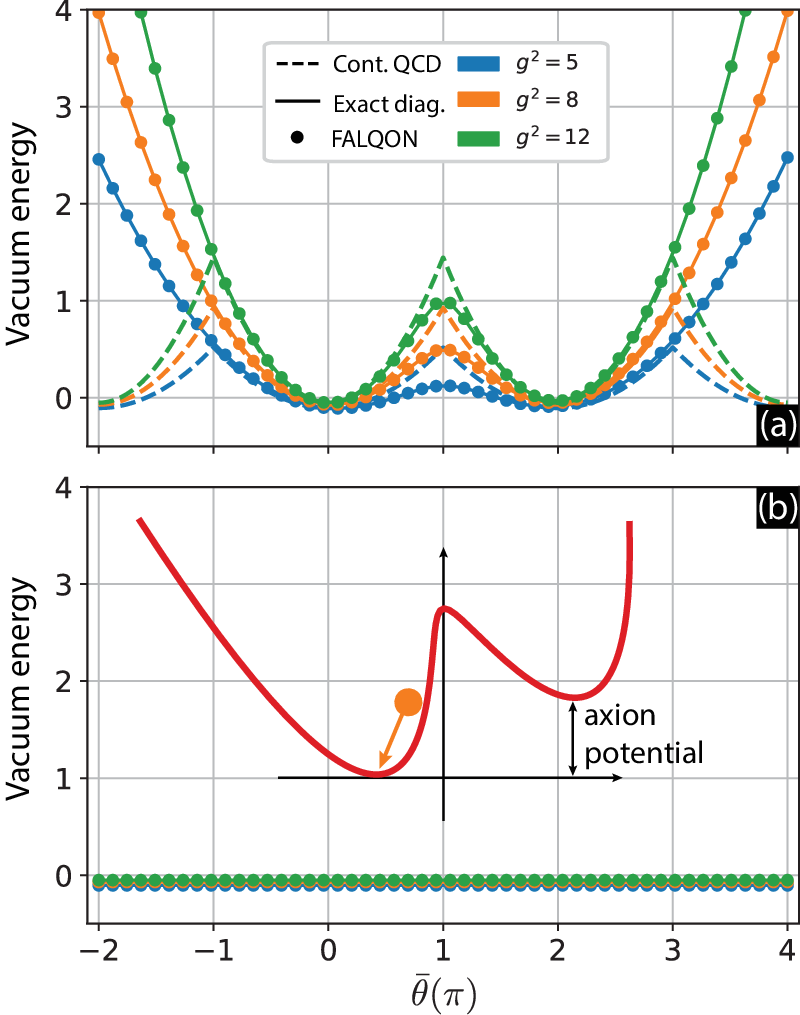}
    \caption{\textbf{Vacuum energy of the two-site lattice Schwinger model with and without a dynamical axion.}
(a) Vacuum energy $E_{\rm vac}(\bar{\theta})$ as a function of $\bar{\theta}$ for three representative parameter sets. Dashed curves show the continuum QCD prediction, solid curves correspond to exact diagonalization, and circular markers are results obtained from the FALQON algorithm. 
(b) Vacuum energy after coupling to a dynamical axion field. \textit{Inset:} Schematic axion-induced effective potential illustrating the relaxation of $\theta_{\rm eff}$ to its global minimum.
}
    \label{fig1}
\end{figure}

The axion Hamiltonian density follows from the canonical momentum $\pi_a=\dot a$,
\begin{align}
\mathcal{H}_a
=
\frac{1}{2}\pi_a^2
+\frac{1}{2}(\partial_x a)^2
+V(\theta_{\rm eff}),
\
H_a=\int dx\,\mathcal{H}_a.
\end{align}
In our $(1+1)$D Hamiltonian, the axion couples through the shifted $\bar{\theta}$ term,
\begin{align}
H_{\bar{\theta}+a}
=
\int dx\,\frac{g^2}{2}
\left(
E(x)-\frac{\theta_{\rm eff}(x)}{2\pi}
\right)^2.
\end{align}
Combining all contributions, the full Hamiltonian is
\begin{align}
H_{\rm total}
=
H_m+H_{\rm hop}+H_{\bar{\theta}+a}+H_a+H_G .
\end{align}
Minimization of the effective potential drives $\theta_{\rm eff}$ to $0$ (mod $2\pi$), i.e., $\langle a\rangle \simeq -f_a\bar{\theta}$, dynamically cancelling the $CP$-violating phase and realizing the PQ mechanism.

%\subsection{Numerical simulation}
\subsection{Numerical simulation}
We compute the ground state of the two-site Schwinger-model Hamiltonian in its qubit representation using the feedback-based FALQON algorithm~\cite{PhysRevLett.129.250502}. FALQON updates control parameters in real time to enforce a strictly monotonic decrease of the energy expectation value, enabling ground-state preparation without a classical optimization loop (see App.~\ref{appB}).

We probe confinement and $CP$-violating effects across different parameter regimes. Specifically, we fix $w=1$, $m=0.01$, and $\lambda=10^6$, and vary the gauge coupling as $g^{2}=5,8,$ and $12$. 
For the axion sector, we choose $m_a=g^{2}$ and $f_a=2$, matching the gauge-field energy scale.

Figure~\ref{fig1}(a) shows the vacuum energy $E_{\rm vac}(\bar{\theta})$ as a function of $\bar{\theta}$ (in units of $\pi$) in the absence of an axion, governed by Eq.~\eqref{eq:Htot}. In continuum QCD, the physical vacuum energy is obtained by minimizing over topological sectors,
\begin{align}
E_{\rm vac}(\bar{\theta})=\min_{n\in\mathbb{Z}}E_n(\bar{\theta}),\
E_n(\bar{\theta})=\frac{g^{2}}{2}\left(n+\frac{\bar{\theta}}{2\pi}\right)^{2},
\end{align}
yielding a $2\pi$-periodic structure with degenerate minima at $\bar{\theta}=0\ (\mathrm{mod}\ 2\pi)$, as indicated by the dashed curves in Fig.~\ref{fig1}(a), where we have shifted its energy to match with the two-site lattice model.

In the two-site lattice model, the ideal $\bar{\theta}$-periodic structure is distorted by reduced dimensionality, lattice discretization, and finite-volume effects. As a result, only the minima at $\bar{\theta}=0$ and $2\pi$ are clearly visible, while intermediate branches are lifted. Both exact diagonalization and FALQON confirm this behavior, as shown by the solid curves and circular markers. 

Figure~\ref{fig1}(b) illustrates the effect of coupling the system to a dynamical axion field. Starting from a generic initial $\bar{\theta}$, the axion relaxes toward the minimum of the combined potential, driving the effective angle $\theta_{\rm eff}=\bar{\theta}+a/f_a$ to zero. As a result, the relaxed vacuum energy becomes independent of $\bar{\theta}$, demonstrating the dynamical cancellation of $CP$ violation. Despite lattice and dimensional artifacts inherent to the $(1+1)$D setting, these results confirm that the PQ mechanism remains operative at the Hamiltonian level.

\subsection{Extended lattice: 4 sites and 3 links (7 qubits)}
\begin{figure}[t]
    \centering
    \includegraphics[width=\linewidth]{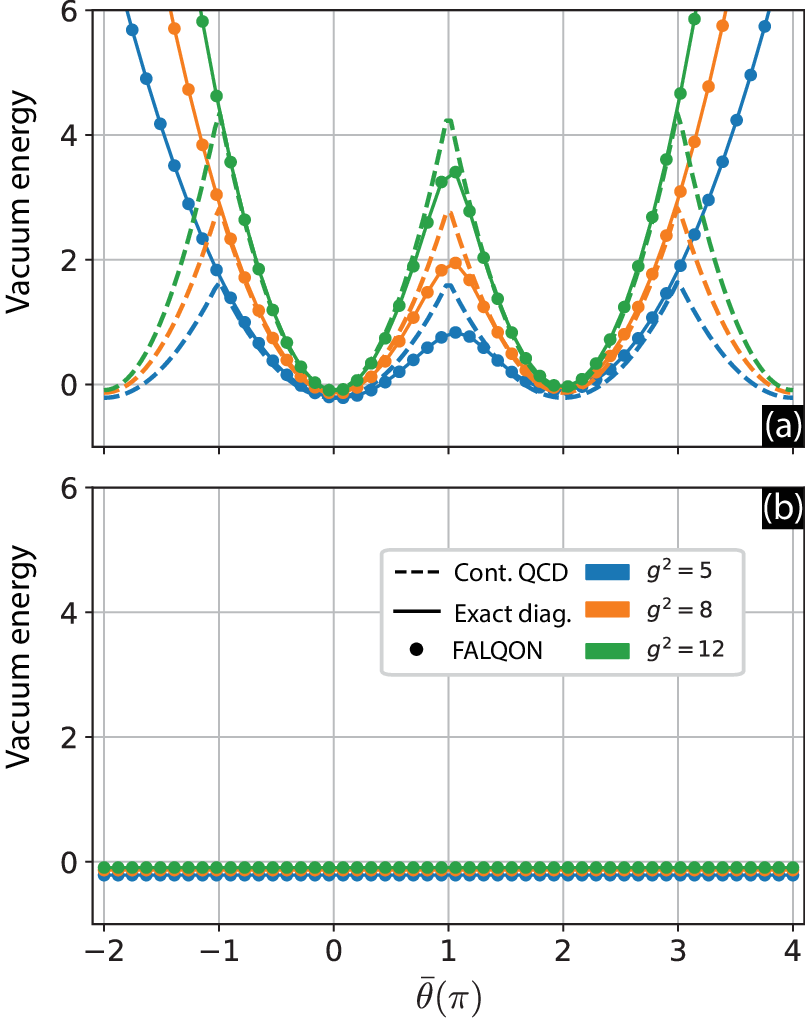}
    \caption{\textbf{Vacuum energy of the four-site lattice Schwinger model with and without a dynamical axion.}
Vacuum energy $E_{\rm vac}(\bar{\theta})$ are plotted for three representative parameter regimes, shown using the same conventions as Fig.~\ref{fig1}: (a) without a dynamical axion and (b) with a dynamical axion.}
    \label{fig2}
\end{figure}

Figure~\ref{fig2} presents the vacuum-energy landscape $E_{\rm vac}(\bar{\theta})$ of the (1+1)D Schwinger model on a four-site lattice with three gauge links. The explicit Hamiltonian, including mass, gauge-assisted hopping, electric-field energy, and Gauss-law terms, is given in App.~\ref{appC}.

Figure~\ref{fig2}(a) shows the case without a dynamical axion. As in the two-site case of Fig.~\ref{fig1}(a), the simulated $E_{\rm vac}(\bar{\theta})$ exhibits minima at $\bar{\theta}=0$ and $2\pi$, while the full $2\pi$-periodic structure is not completely resolved. 
Figure~\ref{fig2}(b) shows the same model coupled to a dynamical axion. As in Fig.~\ref{fig1}(b), the axion shifts the effective angle as $\bar{\theta}\to\theta_{\rm eff}=\bar{\theta}+a/f_a$ and generates a restoring potential. After relaxation, $E_{\rm vac}(\bar{\theta})$ becomes independent of $\bar{\theta}$, with the global minimum fixed at $\theta_{\rm eff}=0$ for all parameter regimes. This demonstrates that the PQ mechanism remains effective despite finite-volume effects, yielding a $CP$-symmetric vacuum in the seven-qubit lattice realization.

\section{Discussion}
The present simulations focus on few-link systems to enable exact benchmarking of the lattice mapping and feedback dynamics. This restriction is practical rather than conceptual. For an $N$-link chain, the quantum link Hamiltonian can be extended straightforwardly, with nearest-neighbour gauge-matter couplings and local electric-field terms defined at each site. The number of required qubits scales linearly with $N$, leading to polynomial growth in computational complexity. 

More generally, the construction can be extended to $(d+1)$-dimensional lattice gauge theories, and to models with multiple fermion flavors. Consider a $d$-dimensional hypercubic lattice with sizes $L_i$ and open boundary conditions, giving $N_s=\prod_{i=1}^{d} L_i$ sites. The number of gauge links is
\(
N_{\mathrm{links}}=\sum_{i=1}^{d}(L_i-1)\prod_{j\neq i}L_j .
\)
Each link requires $N_{\rm tr}=\lceil \log_2 d_{\mathrm{tr}}\rceil$ qubits for a truncating dimension $d_{\mathrm{tr}}$. Thus, the gauge sector uses $N_{\mathrm{links}}N_{\rm tr}$ qubits. The matter sector requires $N_fN_s$ qubits for $N_f$ fermion flavors, giving
\(
N_{\mathrm{tot}} = N_fN_s + N_{\mathrm{links}}N_{\rm tr} .
\)
For example, a $(2+1)$D with $2\times2$ lattice gives $N_s=4$ and $N_{\mathrm{links}}=4$. With $d_{\mathrm{tr}}=4$ gives $N_{\rm tr}=2$, and with $N_f = 2$, the total system size is $16$ qubits.
Because the Hamiltonian contains only local on-site and nearest-neighbor gauge-matter interactions, the number of terms grows linearly with lattice size, allowing extension to larger systems as quantum hardware improves.

\section{Conclusion}
Starting from the QCD Lagrangian with a $CP$-violating $\bar{\theta}$ term, we derived the corresponding Hamiltonian and analyzed its impact on the gauge sector. By reducing the theory to a $(1+1)$D Schwinger-model analogue, we obtained a minimal framework that preserves essential nonperturbative features, including topological vacuum structure and $\bar{\theta}$ dependence. Using Jordan-Wigner and quantum-link mappings, the model was cast as an explicit Pauli Hamiltonian suitable for digital quantum simulation. The ground state was prepared using the FALQON algorithm, yielding a vacuum energy consistent with explicit $CP$ violation. Upon coupling to a dynamical axion field, the vacuum energy relaxes to a global minimum at $\theta_{\rm eff}=0$, demonstrating dynamical cancellation of $CP$ violation. These results realize the PQ mechanism within a controlled quantum-simulation setting and provide a platform for studying $CP$ violation and its dynamical relaxation in gauge theories.
An extension of the present framework to non-Abelian gauge theories remains an interesting direction for future investigation and may further broaden the applicability of the proposed approach.

\acknowledgments
This work is supported by the Tohoku Initiative for Fostering Global Researchers for Interdisciplinary Sciences (TI-FRIS) of MEXT's Strategic Professional Development Program for Young Researchers and FRIS Creative Interdisciplinary Collaboration Program. We thank L. Duc-Truyen and N. Kitajima for useful discussion.

\section*{Data availability} 
%The data are available from the authors upon reasonable request.
The data that support the findings of this article are openly available at \cite{strongcp}.

\appendix
\section{Jordan-Wigner Transformation}\label{appA}
To simulate the $(1+1)$D lattice Schwinger model on a quantum computer, the Hamiltonian must be expressed in terms of Pauli operators acting on qubits. This requires mapping both fermionic matter fields and gauge-link variables to qubit degrees of freedom. We encode the fermions using the Jordan-Wigner (JW) transformation and represent the gauge links using the quantum link model (QLM), which provides a finite-dimensional truncation of the gauge field.

The staggered fermion operators $\psi_x$ and $\psi_x^\dagger$ obey the canonical anticommutation relations
\[
\{\psi_x,\psi_y^\dagger\}=\delta_{xy}, \qquad \{\psi_x,\psi_y\}=0.
\]
Under the JW transformation, they map to Pauli operators as
\begin{align}
\psi_x = \Big(\prod_{y<x} Z_y\Big) S_x^-, \qquad
\psi_x^\dagger = \Big(\prod_{y<x} Z_y\Big) S_x^+,
\end{align}
where $S_x^\pm=(X_x\mp iY_x)/2$, i.e., $S_x^- = |0\rangle\langle1|$, $S_x^+ = |1\rangle\langle0|$, and the string operator $\prod_{y<x} Z_y$ enforces fermionic parity. For the two-site system $(x=0,1)$, this reduces to
\begin{align}\label{app:eq:psi}
\psi_0 = S_0^-, \ \psi_0^\dagger = S_0^+, \
\psi_1 = Z_0 S_1^-, \ \psi_1^\dagger = Z_0 S_1^+.
\end{align}
Each qubit represents a fermionic site, with $|0\rangle$ and $|1\rangle$ corresponding to empty and occupied states, respectively.

The gauge field on the link connecting sites $0$ and $1$ is encoded in a third qubit, labeled $\ell$. In the QLM formulation, the electric field and link operators are
\begin{align}\label{app:eq:EU}
E_\ell \equiv E_{01} = \frac{I - Z_\ell}{2}, \
U_\ell \equiv U_{01} = \frac{X_\ell + iY_\ell}{2}.
\end{align}
Here $E_\ell$ measures the quantized electric flux, while $U_\ell$ and $U_\ell^\dagger$ act as raising and lowering operators satisfying $[E_\ell,U_\ell]=U_\ell$ and $[E_\ell,U_\ell^\dagger]=-U_\ell^\dagger$. The eigenvalues of $E_\ell$ correspond to discrete electric-field strengths $\mathcal{E}\in\{0,1\}$.

Substituting these mappings into the lattice Hamiltonian yields a fully qubit-based representation. The staggered mass term becomes
\begin{align}
H_m = \frac{m}{2}\big(Z_0 - Z_1\big).
\end{align}
The electric-field energy, including the $CP$-violating $\bar{\theta}$ shift, takes the form
\begin{align}
H_{\bar{\theta}}
&= \frac{g^2}{2}\Big(E_\ell-\frac{\bar{\theta}}{2\pi}\Big)^2 \nonumber\\
&= \frac{g^2}{2}\!\left[\!\left(\frac{\bar{\theta}}{2\pi}\right)^2
- \frac{\bar{\theta}}{2\pi} + \frac{1}{2}\!\right] I_\ell
+ \frac{g^2}{2}\!\left(\frac{\bar{\theta}}{2\pi}-\frac{1}{2}\right) Z_\ell .
\end{align}

The gauge-invariant hopping term reads
\begin{align}
H_{\rm hop}
&= \frac{w}{2}\Big(\psi_0^\dagger U_{01}\psi_1 + \psi_1^\dagger U_{01}^\dagger \psi_0\Big).
\end{align}
Substituting Eqs. (\ref{app:eq:psi}, \ref{app:eq:EU}),  yields 
\begin{align}
\psi_0^\dagger U_{01}\psi_1
&= S_0^+\,U_\ell\,(Z_0 S_1^-)\nonumber\\
&= \frac{1}{8}(X_0-iY_0)(X_\ell+iY_\ell)(X_1+iY_1),
\end{align}
where we used $S_0^+Z_0 = |1\rangle\langle 0|Z_0 = S_0^+$, and similarly
\begin{align}
\psi_1^\dagger U_{01}^\dagger\psi_0
&=(Z_0S_1^+)\,U_\ell^\dagger\,S_0^-\nonumber\\
&= \frac{1}{8}(X_1-iY_1)(X_\ell-iY_\ell)(X_0+iY_0).
\end{align}
Adding the two contributions, we obtain
\begin{align}
H_{\rm hop}
= \frac{w}{8}\Big(
X_0 X_1 X_\ell + Y_0 Y_1 X_\ell
- X_0 Y_1 Y_\ell + Y_0 X_1 Y_\ell
\Big),
\end{align}
describing correlated three-body interactions that move a fermion between sites while simultaneously changing the electric flux, thereby preserving local gauge invariance.

The Gauss law is enforced energetically through a penalty term
\begin{align}
H_G = \lambda \sum_x G_x^2 \end{align}
where \(
G_x = E_{x-1,x} - E_{x,x+1} - \psi_x^\dagger\psi_x + \eta_x
\). Here, we use the open-boundary conditions $E_{-1,0}=E_{1,2}=0$,
$\psi_x^\dagger\psi_x = \frac{1}{2}(I-Z_x)$, and $\eta_x = \frac{1}{2}[1+(-1)^x]$
satisfying $\sum_x\eta_x = 1$.

Combining all contributions, the complete two-site Pauli Hamiltonian is
\begin{align}
H_{\rm total} = H_m + H_{\rm hop} + H_{\bar{\theta}} + H_G.
\end{align}
Despite its minimal size, this model retains the essential structure of the Schwinger mechanism and provides a compact benchmark for near-term quantum simulations of gauge-theory dynamics.

\section{Ground-state preparation via FALQON}
\label{appB}
To prepare the ground state of the lattice Schwinger Hamiltonian, we employ the feedback-based algorithm for quantum optimization (FALQON)~\cite{PhysRevLett.129.250502}. FALQON formulates ground-state preparation as a closed-loop quantum control problem, where the system evolves under a time-dependent Hamiltonian
\begin{align}
H(t)=H_C+\beta(t) H_D ,
\end{align}
with $H_C$ the cost (problem) Hamiltonian and $H_D$ a driver that induces state transitions. The feedback field $\beta(t)$ is updated according to
\begin{align}
\beta(t)=-\langle \psi(t) | i[H_D,H_C] | \psi(t) \rangle ,
\end{align}
which guarantees a monotonic decrease of the cost energy,
\begin{align}
\frac{d}{dt}\langle H_C\rangle_t
= - i \big|\langle [H_D,H_C]\rangle_t\big|^2 \le 0 .
\end{align}

In discrete time, the evolution proceeds via $p$ alternating unitaries,
\begin{align}
|\psi_{k+1}\rangle &= e^{-i\Delta t\,\beta_k H_D}\,e^{-i\Delta t\,H_C}\,|\psi_k\rangle ,\\
\beta_{k+1} &= -\langle \psi_k | i[H_D,H_C] | \psi_k \rangle ,
\end{align}
with time step $\Delta t$. In our implementation, we take $H_C=H_{\rm total}$ and choose the global mixer $H_D=\sum_i X_i$. The feedback mechanism automatically adjusts the control amplitudes $\{\beta_k\}$, enabling efficient and robust convergence to the ground state without a classical optimization loop.

\section{Multi-qubit gauge link model}\label{appD}
To systematically enlarge the gauge-field Hilbert space while remaining within a qubit-based framework, we encode each lattice link using $N_\ell$ qubits. This approach extends the minimal single-qubit quantum link truncation to a link space of dimension $d=2^{N_\ell}$, thereby allowing controlled increases in the gauge-field resolution.

We encode the electric field on a link using an $N_\ell$-qubit register. 
The electric-field operator is defined as the binary-weighted sum
\begin{align}
E = \sum_{j=0}^{N_\ell-1} 2^j\, E_j,
\qquad
E_j = \frac{I - Z_j}{2},
\label{eq:E_Nqubit}
\end{align}
where each $E_j$ has eigenvalues $0$ and $1$. 
On a computational basis state $|n_{N_\ell-1}\cdots n_0\rangle$ with $n_j\in\{0,1\}$,
\begin{align}
E\,|n_{N_\ell-1}\cdots n_0\rangle
=
\mathcal{E}\,|n_{N_\ell-1}\cdots n_0\rangle,
\end{align}
where $\mathcal{E}=\sum_{j=0}^{N_\ell-1}2^j n_j$, 
i.e.,  $\mathcal{E}\in\{0,1,\dots,2^{N_\ell}-1\}$ and the link Hilbert-space dimension is $d=2^{N_\ell}$.

In this ``integer" representation, the computational basis $|n_{N_\ell-1}\cdots n_1 n_0\rangle$ reduce to the number basis $|e\rangle$,
such that $E|e\rangle = e|e\rangle$ with $e\in\{0,\dots,d-1\}$.

To implement the gauge-matter hopping term, we introduce the truncated raising operator
\begin{align}
U = \sum_{e=0}^{d-2} |e+1\rangle\langle e|,
\qquad
U^\dagger = \sum_{e=0}^{d-2} |e\rangle\langle e+1|,
\label{eq:U_shift}
\end{align}
which acts as
\begin{align}
U|e\rangle =
\begin{cases}
|e+1\rangle, & e<d-1,\\
0, & e=d-1 .
\end{cases}
\end{align}
This operator is a finite-dimensional truncation of the $U(1)$ link operator.

\begin{figure}[t]
    \centering
    \includegraphics[width=\linewidth]{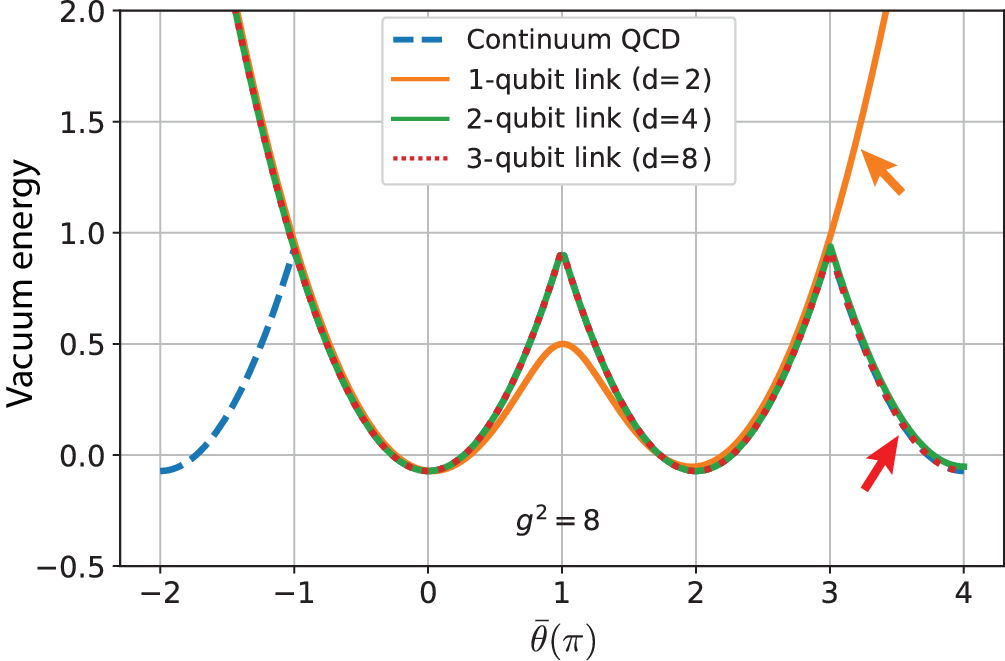}
    \caption{\textbf{Vacuum energy of the two-site lattice Schwinger model as a function of $\bar{\theta}$ for different gauge-link truncations.}
The coupling is fixed at $g^2 = 8$. The dashed blue curve shows the continuum QCD result, while the solid and dotted curves correspond to 1-qubit ($d=2$), 2-qubit ($d=4$), and 3-qubit ($d=8$) gauge-link truncations.
    }
    \label{fig3}
\end{figure}

Figure~\ref{fig3} presents the vacuum energy as a function of $\bar{\theta}$, following the same setup as in Fig.~\ref{fig1}. The numerical results are obtained from the exact diagonal the Hamiltonian. Here, the coupling is fixed to $g^2 = 8$, and we compare different truncation levels of the gauge-link Hilbert space.

For the 1-qubit gauge-link truncation ($d=2$), the lattice results reproduce the continuum QCD behavior only within a limited range of $\bar{\theta}$, approximately $\bar{\theta} \in [-\pi, 3\pi]$, as indicated by the orange arrow. In particular, noticeable deviations appear around $\bar{\theta} = \pi$, where the truncated model fails to accurately capture the continuum behavior.

In contrast, increasing the gauge-link truncation to 2 qubits ($d=4$) and 3 qubits ($d=8$) significantly improves the agreement with the continuum result. The vacuum energy matches well at $\bar{\theta} = \pi$ and the region of agreement also extends further, as highlighted by the red arrow. The 2-qubit and 3-qubit cases are nearly indistinguishable on the scale shown.

These results indicate that enlarging the gauge-link Hilbert space systematically improves the accuracy of the lattice approximation and broadens the range of $\bar\theta$ over which continuum behavior is faithfully reproduced. Increasing the link dimension effectively reduces truncation effects in the electric-field sector, allowing the lattice model to more closely mimic the structure of the underlying gauge theory.

This improvement can be understood in relation to the Kogut-Susskind (KS) limit of lattice gauge theories. In the Hamiltonian formulation, the KS limit corresponds to the strong-coupling or continuum limit, where the lattice description converges to the underlying quantum field theory. In practical quantum simulations with finite-dimensional quantum links, the gauge-field Hilbert space is necessarily truncated. Increasing the link dimension does not by itself realize the full continuum limit, but it brings the system closer to the KS regime by reducing discretization and truncation effects. Thus, enlarging the link Hilbert space provides a controlled approach toward recovering continuum physics within finite resources, at least in the small regime of $\bar\theta$.

\section{Four sites and three links (7 qubits) model}
\label{appC}
We consider a four-site open chain with three gauge links. Matter sites are labeled
$x=0,1,2,3$, and links $\ell=4,5,6$ connect neighboring pairs $(x,x{+}1)$. In the
quantum link formulation, the gauge-field operators are
\begin{align}
E_\ell=\frac{I-Z_\ell}{2}, \qquad
U_\ell=\frac{1}{2}\big(X_\ell+iY_\ell\big).
\end{align}
Using the Jordan-Wigner mapping, we assign
\begin{align}
\text{matter qubits: } & q_0,q_1,q_2,q_3 \leftrightarrow x=0,1,2,3, \\
\text{link qubits: } & q_4,q_5,q_6 \leftrightarrow \ell=4,5,6 .
\end{align}
The total Hamiltonian takes the form
\begin{align}
H = H_m + H_{\rm hop} + H_{\bar{\theta}} + H_G + \mathrm{const.}
\end{align}
The staggered mass term is
\begin{align}
H_m = \frac{m}{2}\big(Z_{q_0}-Z_{q_1}+Z_{q_2}-Z_{q_3}\big).
\end{align}
The electric-field energy, including the $\bar{\theta}$ shift, is
\begin{align}
H_{\bar{\theta}}
= \frac{g^2}{2}\sum_{\ell=4}^{6}\!\left(E_\ell-\frac{\bar{\theta}}{2\pi}\right)^2,
\qquad
E_\ell=\frac{I-Z_{q_\ell}}{2}.
\end{align}

The gauge-invariant hopping term reads
\begin{align}
H_{\rm hop}
&= \frac{w}{2}\sum_{x=0}^{2}\!\left(\psi_x^\dagger U_{x,x+1} \psi_{x+1} + \mathrm{h.c.}\right) \nonumber\\
&= \frac{w}{8}\sum_{x=0}^{2}\!\Big(
X_{q_x}X_{q_{x+1}}X_{q_\ell}
+ Y_{q_x}Y_{q_{x+1}}X_{q_\ell} \nonumber\\
&\hspace{1.2cm}
- X_{q_x}Y_{q_{x+1}}Y_{q_\ell}
+ Y_{q_x}X_{q_{x+1}}Y_{q_\ell}
\Big),
\end{align}
where $\ell=x+4$ labels the link connecting sites $(x,x{+}1)$.

The Gauss law is enforced at each site as
\(
G_x = E_{x-1,x}-E_{x,x+1}-\psi_x^\dagger\psi_x+\eta_x, \)
where 
\(\psi_x^\dagger\psi_x = \frac{1}{2}(I-Z_{q_x} )\) 
and 
\(\eta_x = \tfrac{1}{2}\big[1+(-1)^x\big],
\)
with open-boundary conditions $E_{-1,0}=E_{3,4}=0$. Gauge invariance is imposed through a penalty term
\begin{align}
H_G = \lambda \sum_{x=0}^{3} G_x^2,
\qquad
\lambda \gg \max(m,w,g^2).
\end{align}

\bibliography{refs}

\begin{thebibliography}{36}%
\makeatletter
\providecommand \@ifxundefined [1]{%
 \@ifx{#1\undefined}
}%
\providecommand \@ifnum [1]{%
 \ifnum #1\expandafter \@firstoftwo
 \else \expandafter \@secondoftwo
 \fi
}%
\providecommand \@ifx [1]{%
 \ifx #1\expandafter \@firstoftwo
 \else \expandafter \@secondoftwo
 \fi
}%
\providecommand \natexlab [1]{#1}%
\providecommand \enquote  [1]{``#1''}%
\providecommand \bibnamefont  [1]{#1}%
\providecommand \bibfnamefont [1]{#1}%
\providecommand \citenamefont [1]{#1}%
\providecommand \href@noop [0]{\@secondoftwo}%
\providecommand \href [0]{\begingroup \@sanitize@url \@href}%
\providecommand \@href[1]{\@@startlink{#1}\@@href}%
\providecommand \@@href[1]{\endgroup#1\@@endlink}%
\providecommand \@sanitize@url [0]{\catcode `\\12\catcode `\$12\catcode
  `\&12\catcode `\#12\catcode `\^12\catcode `\_12\catcode `\%12\relax}%
\providecommand \@@startlink[1]{}%
\providecommand \@@endlink[0]{}%
\providecommand \url  [0]{\begingroup\@sanitize@url \@url }%
\providecommand \@url [1]{\endgroup\@href {#1}{\urlprefix }}%
\providecommand \urlprefix  [0]{URL }%
\providecommand \Eprint [0]{\href }%
\providecommand \doibase [0]{https://doi.org/}%
\providecommand \selectlanguage [0]{\@gobble}%
\providecommand \bibinfo  [0]{\@secondoftwo}%
\providecommand \bibfield  [0]{\@secondoftwo}%
\providecommand \translation [1]{[#1]}%
\providecommand \BibitemOpen [0]{}%
\providecommand \bibitemStop [0]{}%
\providecommand \bibitemNoStop [0]{.\EOS\space}%
\providecommand \EOS [0]{\spacefactor3000\relax}%
\providecommand \BibitemShut  [1]{\csname bibitem#1\endcsname}%
\let\auto@bib@innerbib\@empty
%</preamble>
\bibitem [{\citenamefont
  {'t~Hooft}(2000)}]{https://doi.org/10.1002/andp.200051211-1210}%
  \BibitemOpen
  \bibfield  {author} {\bibinfo {author} {\bibfnamefont {G.}~\bibnamefont
  {'t~Hooft}},\ }\bibfield  {title} {\bibinfo {title} {Quantum
  chromodynamics},\ }\href
  {https://doi.org/https://doi.org/10.1002/andp.200051211-1210} {\bibfield
  {journal} {\bibinfo  {journal} {Annalen der Physik}\ }\textbf {\bibinfo
  {volume} {512}},\ \bibinfo {pages} {925} (\bibinfo {year}
  {2000})}\BibitemShut {NoStop}%
\bibitem [{\citenamefont {Gross}\ \emph {et~al.}(2023)\citenamefont {Gross},
  \citenamefont {Klempt}, \citenamefont {Brodsky}, \citenamefont {Buras},
  \citenamefont {Burkert}, \citenamefont {Heinrich}, \citenamefont {Jakobs},
  \citenamefont {Meyer}, \citenamefont {Orginos}, \citenamefont {Strickland},
  \citenamefont {Stachel}, \citenamefont {Zanderighi}, \citenamefont
  {Brambilla}, \citenamefont {Braun-Munzinger}, \citenamefont {Britzger},
  \citenamefont {Capstick}, \citenamefont {Cohen}, \citenamefont {Crede},
  \citenamefont {Constantinou}, \citenamefont {Davies}, \citenamefont
  {Del~Debbio}, \citenamefont {Denig}, \citenamefont {DeTar}, \citenamefont
  {Deur}, \citenamefont {Dokshitzer}, \citenamefont {Dosch}, \citenamefont
  {Dudek}, \citenamefont {Dunford}, \citenamefont {Epelbaum}, \citenamefont
  {Escobedo}, \citenamefont {Fritzsch}, \citenamefont {Fukushima},
  \citenamefont {Gambino}, \citenamefont {Gillberg}, \citenamefont {Gottlieb},
  \citenamefont {Grafstrom}, \citenamefont {Grazzini}, \citenamefont {Grube},
  \citenamefont {Guskov}, \citenamefont {Iijima}, \citenamefont {Ji},
  \citenamefont {Karsch}, \citenamefont {Kluth}, \citenamefont {Kogut},
  \citenamefont {Krauss}, \citenamefont {Kumano}, \citenamefont {Leinweber},
  \citenamefont {Leutwyler}, \citenamefont {Li}, \citenamefont {Li},
  \citenamefont {Malaescu}, \citenamefont {Mariotti}, \citenamefont {Maris},
  \citenamefont {Marzani}, \citenamefont {Melnitchouk}, \citenamefont
  {Messchendorp}, \citenamefont {Meyer}, \citenamefont {Mitchell},
  \citenamefont {Mondal}, \citenamefont {Nerling}, \citenamefont {Neubert},
  \citenamefont {Pappagallo}, \citenamefont {Pastore}, \citenamefont
  {Pel{\'a}ez}, \citenamefont {Puckett}, \citenamefont {Qiu}, \citenamefont
  {Rabbertz}, \citenamefont {Ramos}, \citenamefont {Rossi}, \citenamefont
  {Rustamov}, \citenamefont {Sch{\"a}fer}, \citenamefont {Scherer},
  \citenamefont {Schindler}, \citenamefont {Schramm}, \citenamefont {Shifman},
  \citenamefont {Shuryak}, \citenamefont {Sj{\"o}strand}, \citenamefont
  {Sterman}, \citenamefont {Stewart}, \citenamefont {Stroth}, \citenamefont
  {Swanson}, \citenamefont {de~T{\'e}ramond}, \citenamefont {Thoma},
  \citenamefont {Vairo}, \citenamefont {van Dyk}, \citenamefont {Vary},
  \citenamefont {Virto}, \citenamefont {Vos}, \citenamefont {Weiss},
  \citenamefont {Wobisch}, \citenamefont {Wu}, \citenamefont {Young},
  \citenamefont {Yuan}, \citenamefont {Zhao},\ and\ \citenamefont
  {Zhou}}]{Gross2023}%
  \BibitemOpen
  \bibfield  {author} {\bibinfo {author} {\bibfnamefont {F.}~\bibnamefont
  {Gross}}, \bibinfo {author} {\bibfnamefont {E.}~\bibnamefont {Klempt}},
  \bibinfo {author} {\bibfnamefont {S.~J.}\ \bibnamefont {Brodsky}}, \bibinfo
  {author} {\bibfnamefont {A.~J.}\ \bibnamefont {Buras}}, \bibinfo {author}
  {\bibfnamefont {V.~D.}\ \bibnamefont {Burkert}}, \bibinfo {author}
  {\bibfnamefont {G.}~\bibnamefont {Heinrich}}, \bibinfo {author}
  {\bibfnamefont {K.}~\bibnamefont {Jakobs}}, \bibinfo {author} {\bibfnamefont
  {C.~A.}\ \bibnamefont {Meyer}}, \bibinfo {author} {\bibfnamefont
  {K.}~\bibnamefont {Orginos}}, \bibinfo {author} {\bibfnamefont
  {M.}~\bibnamefont {Strickland}}, \bibinfo {author} {\bibfnamefont
  {J.}~\bibnamefont {Stachel}}, \bibinfo {author} {\bibfnamefont
  {G.}~\bibnamefont {Zanderighi}}, \bibinfo {author} {\bibfnamefont
  {N.}~\bibnamefont {Brambilla}}, \bibinfo {author} {\bibfnamefont
  {P.}~\bibnamefont {Braun-Munzinger}}, \bibinfo {author} {\bibfnamefont
  {D.}~\bibnamefont {Britzger}}, \bibinfo {author} {\bibfnamefont
  {S.}~\bibnamefont {Capstick}}, \bibinfo {author} {\bibfnamefont
  {T.}~\bibnamefont {Cohen}}, \bibinfo {author} {\bibfnamefont
  {V.}~\bibnamefont {Crede}}, \bibinfo {author} {\bibfnamefont
  {M.}~\bibnamefont {Constantinou}}, \bibinfo {author} {\bibfnamefont
  {C.}~\bibnamefont {Davies}}, \bibinfo {author} {\bibfnamefont
  {L.}~\bibnamefont {Del~Debbio}}, \bibinfo {author} {\bibfnamefont
  {A.}~\bibnamefont {Denig}}, \bibinfo {author} {\bibfnamefont
  {C.}~\bibnamefont {DeTar}}, \bibinfo {author} {\bibfnamefont
  {A.}~\bibnamefont {Deur}}, \bibinfo {author} {\bibfnamefont {Y.}~\bibnamefont
  {Dokshitzer}}, \bibinfo {author} {\bibfnamefont {H.~G.}\ \bibnamefont
  {Dosch}}, \bibinfo {author} {\bibfnamefont {J.}~\bibnamefont {Dudek}},
  \bibinfo {author} {\bibfnamefont {M.}~\bibnamefont {Dunford}}, \bibinfo
  {author} {\bibfnamefont {E.}~\bibnamefont {Epelbaum}}, \bibinfo {author}
  {\bibfnamefont {M.~A.}\ \bibnamefont {Escobedo}}, \bibinfo {author}
  {\bibfnamefont {H.}~\bibnamefont {Fritzsch}}, \bibinfo {author}
  {\bibfnamefont {K.}~\bibnamefont {Fukushima}}, \bibinfo {author}
  {\bibfnamefont {P.}~\bibnamefont {Gambino}}, \bibinfo {author} {\bibfnamefont
  {D.}~\bibnamefont {Gillberg}}, \bibinfo {author} {\bibfnamefont
  {S.}~\bibnamefont {Gottlieb}}, \bibinfo {author} {\bibfnamefont
  {P.}~\bibnamefont {Grafstrom}}, \bibinfo {author} {\bibfnamefont
  {M.}~\bibnamefont {Grazzini}}, \bibinfo {author} {\bibfnamefont
  {B.}~\bibnamefont {Grube}}, \bibinfo {author} {\bibfnamefont
  {A.}~\bibnamefont {Guskov}}, \bibinfo {author} {\bibfnamefont
  {T.}~\bibnamefont {Iijima}}, \bibinfo {author} {\bibfnamefont
  {X.}~\bibnamefont {Ji}}, \bibinfo {author} {\bibfnamefont {F.}~\bibnamefont
  {Karsch}}, \bibinfo {author} {\bibfnamefont {S.}~\bibnamefont {Kluth}},
  \bibinfo {author} {\bibfnamefont {J.~B.}\ \bibnamefont {Kogut}}, \bibinfo
  {author} {\bibfnamefont {F.}~\bibnamefont {Krauss}}, \bibinfo {author}
  {\bibfnamefont {S.}~\bibnamefont {Kumano}}, \bibinfo {author} {\bibfnamefont
  {D.}~\bibnamefont {Leinweber}}, \bibinfo {author} {\bibfnamefont
  {H.}~\bibnamefont {Leutwyler}}, \bibinfo {author} {\bibfnamefont {H.-B.}\
  \bibnamefont {Li}}, \bibinfo {author} {\bibfnamefont {Y.}~\bibnamefont {Li}},
  \bibinfo {author} {\bibfnamefont {B.}~\bibnamefont {Malaescu}}, \bibinfo
  {author} {\bibfnamefont {C.}~\bibnamefont {Mariotti}}, \bibinfo {author}
  {\bibfnamefont {P.}~\bibnamefont {Maris}}, \bibinfo {author} {\bibfnamefont
  {S.}~\bibnamefont {Marzani}}, \bibinfo {author} {\bibfnamefont
  {W.}~\bibnamefont {Melnitchouk}}, \bibinfo {author} {\bibfnamefont
  {J.}~\bibnamefont {Messchendorp}}, \bibinfo {author} {\bibfnamefont
  {H.}~\bibnamefont {Meyer}}, \bibinfo {author} {\bibfnamefont {R.~E.}\
  \bibnamefont {Mitchell}}, \bibinfo {author} {\bibfnamefont {C.}~\bibnamefont
  {Mondal}}, \bibinfo {author} {\bibfnamefont {F.}~\bibnamefont {Nerling}},
  \bibinfo {author} {\bibfnamefont {S.}~\bibnamefont {Neubert}}, \bibinfo
  {author} {\bibfnamefont {M.}~\bibnamefont {Pappagallo}}, \bibinfo {author}
  {\bibfnamefont {S.}~\bibnamefont {Pastore}}, \bibinfo {author} {\bibfnamefont
  {J.~R.}\ \bibnamefont {Pel{\'a}ez}}, \bibinfo {author} {\bibfnamefont
  {A.}~\bibnamefont {Puckett}}, \bibinfo {author} {\bibfnamefont
  {J.}~\bibnamefont {Qiu}}, \bibinfo {author} {\bibfnamefont {K.}~\bibnamefont
  {Rabbertz}}, \bibinfo {author} {\bibfnamefont {A.}~\bibnamefont {Ramos}},
  \bibinfo {author} {\bibfnamefont {P.}~\bibnamefont {Rossi}}, \bibinfo
  {author} {\bibfnamefont {A.}~\bibnamefont {Rustamov}}, \bibinfo {author}
  {\bibfnamefont {A.}~\bibnamefont {Sch{\"a}fer}}, \bibinfo {author}
  {\bibfnamefont {S.}~\bibnamefont {Scherer}}, \bibinfo {author} {\bibfnamefont
  {M.}~\bibnamefont {Schindler}}, \bibinfo {author} {\bibfnamefont
  {S.}~\bibnamefont {Schramm}}, \bibinfo {author} {\bibfnamefont
  {M.}~\bibnamefont {Shifman}}, \bibinfo {author} {\bibfnamefont
  {E.}~\bibnamefont {Shuryak}}, \bibinfo {author} {\bibfnamefont
  {T.}~\bibnamefont {Sj{\"o}strand}}, \bibinfo {author} {\bibfnamefont
  {G.}~\bibnamefont {Sterman}}, \bibinfo {author} {\bibfnamefont {I.~W.}\
  \bibnamefont {Stewart}}, \bibinfo {author} {\bibfnamefont {J.}~\bibnamefont
  {Stroth}}, \bibinfo {author} {\bibfnamefont {E.}~\bibnamefont {Swanson}},
  \bibinfo {author} {\bibfnamefont {G.~F.}\ \bibnamefont {de~T{\'e}ramond}},
  \bibinfo {author} {\bibfnamefont {U.}~\bibnamefont {Thoma}}, \bibinfo
  {author} {\bibfnamefont {A.}~\bibnamefont {Vairo}}, \bibinfo {author}
  {\bibfnamefont {D.}~\bibnamefont {van Dyk}}, \bibinfo {author} {\bibfnamefont
  {J.}~\bibnamefont {Vary}}, \bibinfo {author} {\bibfnamefont {J.}~\bibnamefont
  {Virto}}, \bibinfo {author} {\bibfnamefont {M.}~\bibnamefont {Vos}}, \bibinfo
  {author} {\bibfnamefont {C.}~\bibnamefont {Weiss}}, \bibinfo {author}
  {\bibfnamefont {M.}~\bibnamefont {Wobisch}}, \bibinfo {author} {\bibfnamefont
  {S.~L.}\ \bibnamefont {Wu}}, \bibinfo {author} {\bibfnamefont
  {C.}~\bibnamefont {Young}}, \bibinfo {author} {\bibfnamefont
  {F.}~\bibnamefont {Yuan}}, \bibinfo {author} {\bibfnamefont {X.}~\bibnamefont
  {Zhao}},\ and\ \bibinfo {author} {\bibfnamefont {X.}~\bibnamefont {Zhou}},\
  }\bibfield  {title} {\bibinfo {title} {50 years of quantum chromodynamics},\
  }\href {https://doi.org/10.1140/epjc/s10052-023-11949-2} {\bibfield
  {journal} {\bibinfo  {journal} {The European Physical Journal C}\ }\textbf
  {\bibinfo {volume} {83}},\ \bibinfo {pages} {1125} (\bibinfo {year}
  {2023})}\BibitemShut {NoStop}%
\bibitem [{\citenamefont {Lik\'{e}n\'{e}}\ \emph {et~al.}(2024)\citenamefont
  {Lik\'{e}n\'{e}}, \citenamefont {Ongodo}, \citenamefont {Tsila},
  \citenamefont {Atangana},\ and\ \citenamefont
  {Ben-Bolie}}]{doi:10.1142/S0217732324501943}%
  \BibitemOpen
  \bibfield  {author} {\bibinfo {author} {\bibfnamefont {A.~A.~A.}\
  \bibnamefont {Lik\'{e}n\'{e}}}, \bibinfo {author} {\bibfnamefont {D.~N.}\
  \bibnamefont {Ongodo}}, \bibinfo {author} {\bibfnamefont {P.~M.}\
  \bibnamefont {Tsila}}, \bibinfo {author} {\bibfnamefont {A.}~\bibnamefont
  {Atangana}},\ and\ \bibinfo {author} {\bibfnamefont {G.~H.}\ \bibnamefont
  {Ben-Bolie}},\ }\bibfield  {title} {\bibinfo {title} {Quantum chromodynamics
  lagrangian density and su(3) gauge symmetry: A fractional approach},\ }\href
  {https://doi.org/10.1142/S0217732324501943} {\bibfield  {journal} {\bibinfo
  {journal} {Modern Physics Letters A}\ }\textbf {\bibinfo {volume} {39}},\
  \bibinfo {pages} {2450194} (\bibinfo {year} {2024})}\BibitemShut {NoStop}%
\bibitem [{\citenamefont {'t~Hooft}(1976{\natexlab{a}})}]{PhysRevLett.37.8}%
  \BibitemOpen
  \bibfield  {author} {\bibinfo {author} {\bibfnamefont {G.}~\bibnamefont
  {'t~Hooft}},\ }\bibfield  {title} {\bibinfo {title} {Symmetry breaking
  through bell-jackiw anomalies},\ }\href
  {https://doi.org/10.1103/PhysRevLett.37.8} {\bibfield  {journal} {\bibinfo
  {journal} {Phys. Rev. Lett.}\ }\textbf {\bibinfo {volume} {37}},\ \bibinfo
  {pages} {8} (\bibinfo {year} {1976}{\natexlab{a}})}\BibitemShut {NoStop}%
\bibitem [{\citenamefont {'t~Hooft}(1976{\natexlab{b}})}]{PhysRevD.14.3432}%
  \BibitemOpen
  \bibfield  {author} {\bibinfo {author} {\bibfnamefont {G.}~\bibnamefont
  {'t~Hooft}},\ }\bibfield  {title} {\bibinfo {title} {Computation of the
  quantum effects due to a four-dimensional pseudoparticle},\ }\href
  {https://doi.org/10.1103/PhysRevD.14.3432} {\bibfield  {journal} {\bibinfo
  {journal} {Phys. Rev. D}\ }\textbf {\bibinfo {volume} {14}},\ \bibinfo
  {pages} {3432} (\bibinfo {year} {1976}{\natexlab{b}})}\BibitemShut {NoStop}%
\bibitem [{\citenamefont {Kim}\ and\ \citenamefont
  {Carosi}(2010)}]{RevModPhys.82.557}%
  \BibitemOpen
  \bibfield  {author} {\bibinfo {author} {\bibfnamefont {J.~E.}\ \bibnamefont
  {Kim}}\ and\ \bibinfo {author} {\bibfnamefont {G.}~\bibnamefont {Carosi}},\
  }\bibfield  {title} {\bibinfo {title} {Axions and the strong $cp$ problem},\
  }\href {https://doi.org/10.1103/RevModPhys.82.557} {\bibfield  {journal}
  {\bibinfo  {journal} {Rev. Mod. Phys.}\ }\textbf {\bibinfo {volume} {82}},\
  \bibinfo {pages} {557} (\bibinfo {year} {2010})}\BibitemShut {NoStop}%
\bibitem [{\citenamefont {Crewther}\ \emph {et~al.}(1979)\citenamefont
  {Crewther}, \citenamefont {{Di Vecchia}}, \citenamefont {Veneziano},\ and\
  \citenamefont {Witten}}]{CREWTHER1979123}%
  \BibitemOpen
  \bibfield  {author} {\bibinfo {author} {\bibfnamefont {R.}~\bibnamefont
  {Crewther}}, \bibinfo {author} {\bibfnamefont {P.}~\bibnamefont {{Di
  Vecchia}}}, \bibinfo {author} {\bibfnamefont {G.}~\bibnamefont {Veneziano}},\
  and\ \bibinfo {author} {\bibfnamefont {E.}~\bibnamefont {Witten}},\
  }\bibfield  {title} {\bibinfo {title} {Chiral estimate of the electric dipole
  moment of the neutron in quantum chromodynamics},\ }\href
  {https://doi.org/https://doi.org/10.1016/0370-2693(79)90128-X} {\bibfield
  {journal} {\bibinfo  {journal} {Physics Letters B}\ }\textbf {\bibinfo
  {volume} {88}},\ \bibinfo {pages} {123} (\bibinfo {year} {1979})}\BibitemShut
  {NoStop}%
\bibitem [{\citenamefont {Pospelov}\ and\ \citenamefont
  {Ritz}(2005)}]{POSPELOV2005119}%
  \BibitemOpen
  \bibfield  {author} {\bibinfo {author} {\bibfnamefont {M.}~\bibnamefont
  {Pospelov}}\ and\ \bibinfo {author} {\bibfnamefont {A.}~\bibnamefont
  {Ritz}},\ }\bibfield  {title} {\bibinfo {title} {Electric dipole moments as
  probes of new physics},\ }\href
  {https://doi.org/https://doi.org/10.1016/j.aop.2005.04.002} {\bibfield
  {journal} {\bibinfo  {journal} {Annals of Physics}\ }\textbf {\bibinfo
  {volume} {318}},\ \bibinfo {pages} {119} (\bibinfo {year} {2005})},\ \bibinfo
  {note} {special Issue}\BibitemShut {NoStop}%
\bibitem [{\citenamefont {Baker}\ \emph {et~al.}(2006)\citenamefont {Baker},
  \citenamefont {Doyle}, \citenamefont {Geltenbort}, \citenamefont {Green},
  \citenamefont {van~der Grinten}, \citenamefont {Harris}, \citenamefont
  {Iaydjiev}, \citenamefont {Ivanov}, \citenamefont {May}, \citenamefont
  {Pendlebury}, \citenamefont {Richardson}, \citenamefont {Shiers},\ and\
  \citenamefont {Smith}}]{PhysRevLett.97.131801}%
  \BibitemOpen
  \bibfield  {author} {\bibinfo {author} {\bibfnamefont {C.~A.}\ \bibnamefont
  {Baker}}, \bibinfo {author} {\bibfnamefont {D.~D.}\ \bibnamefont {Doyle}},
  \bibinfo {author} {\bibfnamefont {P.}~\bibnamefont {Geltenbort}}, \bibinfo
  {author} {\bibfnamefont {K.}~\bibnamefont {Green}}, \bibinfo {author}
  {\bibfnamefont {M.~G.~D.}\ \bibnamefont {van~der Grinten}}, \bibinfo {author}
  {\bibfnamefont {P.~G.}\ \bibnamefont {Harris}}, \bibinfo {author}
  {\bibfnamefont {P.}~\bibnamefont {Iaydjiev}}, \bibinfo {author}
  {\bibfnamefont {S.~N.}\ \bibnamefont {Ivanov}}, \bibinfo {author}
  {\bibfnamefont {D.~J.~R.}\ \bibnamefont {May}}, \bibinfo {author}
  {\bibfnamefont {J.~M.}\ \bibnamefont {Pendlebury}}, \bibinfo {author}
  {\bibfnamefont {J.~D.}\ \bibnamefont {Richardson}}, \bibinfo {author}
  {\bibfnamefont {D.}~\bibnamefont {Shiers}},\ and\ \bibinfo {author}
  {\bibfnamefont {K.~F.}\ \bibnamefont {Smith}},\ }\bibfield  {title} {\bibinfo
  {title} {Improved experimental limit on the electric dipole moment of the
  neutron},\ }\href {https://doi.org/10.1103/PhysRevLett.97.131801} {\bibfield
  {journal} {\bibinfo  {journal} {Phys. Rev. Lett.}\ }\textbf {\bibinfo
  {volume} {97}},\ \bibinfo {pages} {131801} (\bibinfo {year}
  {2006})}\BibitemShut {NoStop}%
\bibitem [{\citenamefont {Abel}\ \emph {et~al.}(2020)\citenamefont {Abel},
  \citenamefont {Afach}, \citenamefont {Ayres}, \citenamefont {Baker},
  \citenamefont {Ban}, \citenamefont {Bison}, \citenamefont {Bodek},
  \citenamefont {Bondar}, \citenamefont {Burghoff}, \citenamefont {Chanel},
  \citenamefont {Chowdhuri}, \citenamefont {Chiu}, \citenamefont {Clement},
  \citenamefont {Crawford}, \citenamefont {Daum}, \citenamefont {Emmenegger},
  \citenamefont {Ferraris-Bouchez}, \citenamefont {Fertl}, \citenamefont
  {Flaux}, \citenamefont {Franke}, \citenamefont {Fratangelo}, \citenamefont
  {Geltenbort}, \citenamefont {Green}, \citenamefont {Griffith}, \citenamefont
  {van~der Grinten}, \citenamefont {Gruji\ifmmode~\acute{c}\else \'{c}\fi{}},
  \citenamefont {Harris}, \citenamefont {Hayen}, \citenamefont {Heil},
  \citenamefont {Henneck}, \citenamefont {H\'elaine}, \citenamefont {Hild},
  \citenamefont {Hodge}, \citenamefont {Horras}, \citenamefont {Iaydjiev},
  \citenamefont {Ivanov}, \citenamefont {Kasprzak}, \citenamefont {Kermaidic},
  \citenamefont {Kirch}, \citenamefont {Knecht}, \citenamefont {Knowles},
  \citenamefont {Koch}, \citenamefont {Koss}, \citenamefont {Komposch},
  \citenamefont {Kozela}, \citenamefont {Kraft}, \citenamefont {Krempel},
  \citenamefont {Ku\ifmmode~\acute{z}\else \'{z}\fi{}niak}, \citenamefont
  {Lauss}, \citenamefont {Lefort}, \citenamefont {Lemi\`ere}, \citenamefont
  {Leredde}, \citenamefont {Mohanmurthy}, \citenamefont {Mtchedlishvili},
  \citenamefont {Musgrave}, \citenamefont {Naviliat-Cuncic}, \citenamefont
  {Pais}, \citenamefont {Piegsa}, \citenamefont {Pierre}, \citenamefont
  {Pignol}, \citenamefont {Plonka-Spehr}, \citenamefont {Prashanth},
  \citenamefont {Qu\'em\'ener}, \citenamefont {Rawlik}, \citenamefont
  {Rebreyend}, \citenamefont {Rien\"acker}, \citenamefont {Ries}, \citenamefont
  {Roccia}, \citenamefont {Rogel}, \citenamefont {Rozpedzik}, \citenamefont
  {Schnabel}, \citenamefont {Schmidt-Wellenburg}, \citenamefont {Severijns},
  \citenamefont {Shiers}, \citenamefont {Tavakoli~Dinani}, \citenamefont
  {Thorne}, \citenamefont {Virot}, \citenamefont {Voigt}, \citenamefont {Weis},
  \citenamefont {Wursten}, \citenamefont {Wyszynski}, \citenamefont {Zejma},
  \citenamefont {Zenner},\ and\ \citenamefont
  {Zsigmond}}]{PhysRevLett.124.081803}%
  \BibitemOpen
  \bibfield  {author} {\bibinfo {author} {\bibfnamefont {C.}~\bibnamefont
  {Abel}}, \bibinfo {author} {\bibfnamefont {S.}~\bibnamefont {Afach}},
  \bibinfo {author} {\bibfnamefont {N.~J.}\ \bibnamefont {Ayres}}, \bibinfo
  {author} {\bibfnamefont {C.~A.}\ \bibnamefont {Baker}}, \bibinfo {author}
  {\bibfnamefont {G.}~\bibnamefont {Ban}}, \bibinfo {author} {\bibfnamefont
  {G.}~\bibnamefont {Bison}}, \bibinfo {author} {\bibfnamefont
  {K.}~\bibnamefont {Bodek}}, \bibinfo {author} {\bibfnamefont
  {V.}~\bibnamefont {Bondar}}, \bibinfo {author} {\bibfnamefont
  {M.}~\bibnamefont {Burghoff}}, \bibinfo {author} {\bibfnamefont
  {E.}~\bibnamefont {Chanel}}, \bibinfo {author} {\bibfnamefont
  {Z.}~\bibnamefont {Chowdhuri}}, \bibinfo {author} {\bibfnamefont {P.-J.}\
  \bibnamefont {Chiu}}, \bibinfo {author} {\bibfnamefont {B.}~\bibnamefont
  {Clement}}, \bibinfo {author} {\bibfnamefont {C.~B.}\ \bibnamefont
  {Crawford}}, \bibinfo {author} {\bibfnamefont {M.}~\bibnamefont {Daum}},
  \bibinfo {author} {\bibfnamefont {S.}~\bibnamefont {Emmenegger}}, \bibinfo
  {author} {\bibfnamefont {L.}~\bibnamefont {Ferraris-Bouchez}}, \bibinfo
  {author} {\bibfnamefont {M.}~\bibnamefont {Fertl}}, \bibinfo {author}
  {\bibfnamefont {P.}~\bibnamefont {Flaux}}, \bibinfo {author} {\bibfnamefont
  {B.}~\bibnamefont {Franke}}, \bibinfo {author} {\bibfnamefont
  {A.}~\bibnamefont {Fratangelo}}, \bibinfo {author} {\bibfnamefont
  {P.}~\bibnamefont {Geltenbort}}, \bibinfo {author} {\bibfnamefont
  {K.}~\bibnamefont {Green}}, \bibinfo {author} {\bibfnamefont {W.~C.}\
  \bibnamefont {Griffith}}, \bibinfo {author} {\bibfnamefont {M.}~\bibnamefont
  {van~der Grinten}}, \bibinfo {author} {\bibfnamefont {Z.~D.}\ \bibnamefont
  {Gruji\ifmmode~\acute{c}\else \'{c}\fi{}}}, \bibinfo {author} {\bibfnamefont
  {P.~G.}\ \bibnamefont {Harris}}, \bibinfo {author} {\bibfnamefont
  {L.}~\bibnamefont {Hayen}}, \bibinfo {author} {\bibfnamefont
  {W.}~\bibnamefont {Heil}}, \bibinfo {author} {\bibfnamefont {R.}~\bibnamefont
  {Henneck}}, \bibinfo {author} {\bibfnamefont {V.}~\bibnamefont {H\'elaine}},
  \bibinfo {author} {\bibfnamefont {N.}~\bibnamefont {Hild}}, \bibinfo {author}
  {\bibfnamefont {Z.}~\bibnamefont {Hodge}}, \bibinfo {author} {\bibfnamefont
  {M.}~\bibnamefont {Horras}}, \bibinfo {author} {\bibfnamefont
  {P.}~\bibnamefont {Iaydjiev}}, \bibinfo {author} {\bibfnamefont {S.~N.}\
  \bibnamefont {Ivanov}}, \bibinfo {author} {\bibfnamefont {M.}~\bibnamefont
  {Kasprzak}}, \bibinfo {author} {\bibfnamefont {Y.}~\bibnamefont {Kermaidic}},
  \bibinfo {author} {\bibfnamefont {K.}~\bibnamefont {Kirch}}, \bibinfo
  {author} {\bibfnamefont {A.}~\bibnamefont {Knecht}}, \bibinfo {author}
  {\bibfnamefont {P.}~\bibnamefont {Knowles}}, \bibinfo {author} {\bibfnamefont
  {H.-C.}\ \bibnamefont {Koch}}, \bibinfo {author} {\bibfnamefont {P.~A.}\
  \bibnamefont {Koss}}, \bibinfo {author} {\bibfnamefont {S.}~\bibnamefont
  {Komposch}}, \bibinfo {author} {\bibfnamefont {A.}~\bibnamefont {Kozela}},
  \bibinfo {author} {\bibfnamefont {A.}~\bibnamefont {Kraft}}, \bibinfo
  {author} {\bibfnamefont {J.}~\bibnamefont {Krempel}}, \bibinfo {author}
  {\bibfnamefont {M.}~\bibnamefont {Ku\ifmmode~\acute{z}\else \'{z}\fi{}niak}},
  \bibinfo {author} {\bibfnamefont {B.}~\bibnamefont {Lauss}}, \bibinfo
  {author} {\bibfnamefont {T.}~\bibnamefont {Lefort}}, \bibinfo {author}
  {\bibfnamefont {Y.}~\bibnamefont {Lemi\`ere}}, \bibinfo {author}
  {\bibfnamefont {A.}~\bibnamefont {Leredde}}, \bibinfo {author} {\bibfnamefont
  {P.}~\bibnamefont {Mohanmurthy}}, \bibinfo {author} {\bibfnamefont
  {A.}~\bibnamefont {Mtchedlishvili}}, \bibinfo {author} {\bibfnamefont
  {M.}~\bibnamefont {Musgrave}}, \bibinfo {author} {\bibfnamefont
  {O.}~\bibnamefont {Naviliat-Cuncic}}, \bibinfo {author} {\bibfnamefont
  {D.}~\bibnamefont {Pais}}, \bibinfo {author} {\bibfnamefont {F.~M.}\
  \bibnamefont {Piegsa}}, \bibinfo {author} {\bibfnamefont {E.}~\bibnamefont
  {Pierre}}, \bibinfo {author} {\bibfnamefont {G.}~\bibnamefont {Pignol}},
  \bibinfo {author} {\bibfnamefont {C.}~\bibnamefont {Plonka-Spehr}}, \bibinfo
  {author} {\bibfnamefont {P.~N.}\ \bibnamefont {Prashanth}}, \bibinfo {author}
  {\bibfnamefont {G.}~\bibnamefont {Qu\'em\'ener}}, \bibinfo {author}
  {\bibfnamefont {M.}~\bibnamefont {Rawlik}}, \bibinfo {author} {\bibfnamefont
  {D.}~\bibnamefont {Rebreyend}}, \bibinfo {author} {\bibfnamefont
  {I.}~\bibnamefont {Rien\"acker}}, \bibinfo {author} {\bibfnamefont
  {D.}~\bibnamefont {Ries}}, \bibinfo {author} {\bibfnamefont {S.}~\bibnamefont
  {Roccia}}, \bibinfo {author} {\bibfnamefont {G.}~\bibnamefont {Rogel}},
  \bibinfo {author} {\bibfnamefont {D.}~\bibnamefont {Rozpedzik}}, \bibinfo
  {author} {\bibfnamefont {A.}~\bibnamefont {Schnabel}}, \bibinfo {author}
  {\bibfnamefont {P.}~\bibnamefont {Schmidt-Wellenburg}}, \bibinfo {author}
  {\bibfnamefont {N.}~\bibnamefont {Severijns}}, \bibinfo {author}
  {\bibfnamefont {D.}~\bibnamefont {Shiers}}, \bibinfo {author} {\bibfnamefont
  {R.}~\bibnamefont {Tavakoli~Dinani}}, \bibinfo {author} {\bibfnamefont
  {J.~A.}\ \bibnamefont {Thorne}}, \bibinfo {author} {\bibfnamefont
  {R.}~\bibnamefont {Virot}}, \bibinfo {author} {\bibfnamefont
  {J.}~\bibnamefont {Voigt}}, \bibinfo {author} {\bibfnamefont
  {A.}~\bibnamefont {Weis}}, \bibinfo {author} {\bibfnamefont {E.}~\bibnamefont
  {Wursten}}, \bibinfo {author} {\bibfnamefont {G.}~\bibnamefont {Wyszynski}},
  \bibinfo {author} {\bibfnamefont {J.}~\bibnamefont {Zejma}}, \bibinfo
  {author} {\bibfnamefont {J.}~\bibnamefont {Zenner}},\ and\ \bibinfo {author}
  {\bibfnamefont {G.}~\bibnamefont {Zsigmond}},\ }\bibfield  {title} {\bibinfo
  {title} {Measurement of the permanent electric dipole moment of the
  neutron},\ }\href {https://doi.org/10.1103/PhysRevLett.124.081803} {\bibfield
   {journal} {\bibinfo  {journal} {Phys. Rev. Lett.}\ }\textbf {\bibinfo
  {volume} {124}},\ \bibinfo {pages} {081803} (\bibinfo {year}
  {2020})}\BibitemShut {NoStop}%
\bibitem [{\citenamefont {Peccei}\ and\ \citenamefont
  {Quinn}(1977{\natexlab{a}})}]{PhysRevLett.38.1440}%
  \BibitemOpen
  \bibfield  {author} {\bibinfo {author} {\bibfnamefont {R.~D.}\ \bibnamefont
  {Peccei}}\ and\ \bibinfo {author} {\bibfnamefont {H.~R.}\ \bibnamefont
  {Quinn}},\ }\bibfield  {title} {\bibinfo {title} {$\mathrm{CP}$ conservation
  in the presence of pseudoparticles},\ }\href
  {https://doi.org/10.1103/PhysRevLett.38.1440} {\bibfield  {journal} {\bibinfo
   {journal} {Phys. Rev. Lett.}\ }\textbf {\bibinfo {volume} {38}},\ \bibinfo
  {pages} {1440} (\bibinfo {year} {1977}{\natexlab{a}})}\BibitemShut {NoStop}%
\bibitem [{\citenamefont {Peccei}\ and\ \citenamefont
  {Quinn}(1977{\natexlab{b}})}]{PhysRevD.16.1791}%
  \BibitemOpen
  \bibfield  {author} {\bibinfo {author} {\bibfnamefont {R.~D.}\ \bibnamefont
  {Peccei}}\ and\ \bibinfo {author} {\bibfnamefont {H.~R.}\ \bibnamefont
  {Quinn}},\ }\bibfield  {title} {\bibinfo {title} {Constraints imposed by
  $\mathrm{CP}$ conservation in the presence of pseudoparticles},\ }\href
  {https://doi.org/10.1103/PhysRevD.16.1791} {\bibfield  {journal} {\bibinfo
  {journal} {Phys. Rev. D}\ }\textbf {\bibinfo {volume} {16}},\ \bibinfo
  {pages} {1791} (\bibinfo {year} {1977}{\natexlab{b}})}\BibitemShut {NoStop}%
\bibitem [{\citenamefont {Peccei}(2008)}]{Peccei2008}%
  \BibitemOpen
  \bibfield  {author} {\bibinfo {author} {\bibfnamefont {R.~D.}\ \bibnamefont
  {Peccei}},\ }\bibinfo {title} {The strong cp problem and axions},\ in\ \href
  {https://doi.org/10.1007/978-3-540-73518-2_1} {\emph {\bibinfo {booktitle}
  {Axions: Theory, Cosmology, and Experimental Searches}}},\ \bibinfo {editor}
  {edited by\ \bibinfo {editor} {\bibfnamefont {M.}~\bibnamefont {Kuster}},
  \bibinfo {editor} {\bibfnamefont {G.}~\bibnamefont {Raffelt}},\ and\ \bibinfo
  {editor} {\bibfnamefont {B.}~\bibnamefont {Beltr{\'a}n}}}\ (\bibinfo
  {publisher} {Springer Berlin Heidelberg},\ \bibinfo {address} {Berlin,
  Heidelberg},\ \bibinfo {year} {2008})\ pp.\ \bibinfo {pages}
  {3--17}\BibitemShut {NoStop}%
\bibitem [{\citenamefont {Chadha-Day}\ \emph {et~al.}(2022)\citenamefont
  {Chadha-Day}, \citenamefont {Ellis},\ and\ \citenamefont
  {Marsh}}]{doi:10.1126/sciadv.abj3618}%
  \BibitemOpen
  \bibfield  {author} {\bibinfo {author} {\bibfnamefont {F.}~\bibnamefont
  {Chadha-Day}}, \bibinfo {author} {\bibfnamefont {J.}~\bibnamefont {Ellis}},\
  and\ \bibinfo {author} {\bibfnamefont {D.~J.~E.}\ \bibnamefont {Marsh}},\
  }\bibfield  {title} {\bibinfo {title} {Axion dark matter: What is it and why
  now?},\ }\href {https://doi.org/10.1126/sciadv.abj3618} {\bibfield  {journal}
  {\bibinfo  {journal} {Science Advances}\ }\textbf {\bibinfo {volume} {8}},\
  \bibinfo {pages} {eabj3618} (\bibinfo {year} {2022})},\ \Eprint
  {https://arxiv.org/abs/https://www.science.org/doi/pdf/10.1126/sciadv.abj3618}
  {https://www.science.org/doi/pdf/10.1126/sciadv.abj3618} \BibitemShut
  {NoStop}%
\bibitem [{\citenamefont {Nagata}(2022)}]{NAGATA2022103991}%
  \BibitemOpen
  \bibfield  {author} {\bibinfo {author} {\bibfnamefont {K.}~\bibnamefont
  {Nagata}},\ }\bibfield  {title} {\bibinfo {title} {Finite-density lattice qcd
  and sign problem: Current status and open problems},\ }\href
  {https://doi.org/https://doi.org/10.1016/j.ppnp.2022.103991} {\bibfield
  {journal} {\bibinfo  {journal} {Progress in Particle and Nuclear Physics}\
  }\textbf {\bibinfo {volume} {127}},\ \bibinfo {pages} {103991} (\bibinfo
  {year} {2022})}\BibitemShut {NoStop}%
\bibitem [{\citenamefont {Allés}\ \emph {et~al.}(1996)\citenamefont {Allés},
  \citenamefont {Boyd}, \citenamefont {D'Elia}, \citenamefont {{Di Giacomo}},\
  and\ \citenamefont {Vicari}}]{ALLES1996107}%
  \BibitemOpen
  \bibfield  {author} {\bibinfo {author} {\bibfnamefont {B.}~\bibnamefont
  {Allés}}, \bibinfo {author} {\bibfnamefont {G.}~\bibnamefont {Boyd}},
  \bibinfo {author} {\bibfnamefont {M.}~\bibnamefont {D'Elia}}, \bibinfo
  {author} {\bibfnamefont {A.}~\bibnamefont {{Di Giacomo}}},\ and\ \bibinfo
  {author} {\bibfnamefont {E.}~\bibnamefont {Vicari}},\ }\bibfield  {title}
  {\bibinfo {title} {Hybrid monte carlo and topological modes of full qcd},\
  }\href {https://doi.org/https://doi.org/10.1016/S0370-2693(96)01247-6}
  {\bibfield  {journal} {\bibinfo  {journal} {Physics Letters B}\ }\textbf
  {\bibinfo {volume} {389}},\ \bibinfo {pages} {107} (\bibinfo {year}
  {1996})}\BibitemShut {NoStop}%
\bibitem [{\citenamefont {Bonanno}\ \emph {et~al.}(2024)\citenamefont
  {Bonanno}, \citenamefont {Clemente}, \citenamefont {D'Elia}, \citenamefont
  {Maio},\ and\ \citenamefont {Parente}}]{Bonanno2024}%
  \BibitemOpen
  \bibfield  {author} {\bibinfo {author} {\bibfnamefont {C.}~\bibnamefont
  {Bonanno}}, \bibinfo {author} {\bibfnamefont {G.}~\bibnamefont {Clemente}},
  \bibinfo {author} {\bibfnamefont {M.}~\bibnamefont {D'Elia}}, \bibinfo
  {author} {\bibfnamefont {L.}~\bibnamefont {Maio}},\ and\ \bibinfo {author}
  {\bibfnamefont {L.}~\bibnamefont {Parente}},\ }\bibfield  {title} {\bibinfo
  {title} {Full qcd with milder topological freezing},\ }\href
  {https://doi.org/10.1007/JHEP08(2024)236} {\bibfield  {journal} {\bibinfo
  {journal} {Journal of High Energy Physics}\ }\textbf {\bibinfo {volume}
  {2024}},\ \bibinfo {pages} {236} (\bibinfo {year} {2024})}\BibitemShut
  {NoStop}%
\bibitem [{\citenamefont {Bauer}\ \emph {et~al.}(2023)\citenamefont {Bauer},
  \citenamefont {Davoudi}, \citenamefont {Balantekin}, \citenamefont
  {Bhattacharya}, \citenamefont {Carena}, \citenamefont {de~Jong},
  \citenamefont {Draper}, \citenamefont {El-Khadra}, \citenamefont {Gemelke},
  \citenamefont {Hanada}, \citenamefont {Kharzeev}, \citenamefont {Lamm},
  \citenamefont {Li}, \citenamefont {Liu}, \citenamefont {Lukin}, \citenamefont
  {Meurice}, \citenamefont {Monroe}, \citenamefont {Nachman}, \citenamefont
  {Pagano}, \citenamefont {Preskill}, \citenamefont {Rinaldi}, \citenamefont
  {Roggero}, \citenamefont {Santiago}, \citenamefont {Savage}, \citenamefont
  {Siddiqi}, \citenamefont {Siopsis}, \citenamefont {Van~Zanten}, \citenamefont
  {Wiebe}, \citenamefont {Yamauchi}, \citenamefont {Yeter-Aydeniz},\ and\
  \citenamefont {Zorzetti}}]{PRXQuantum.4.027001}%
  \BibitemOpen
  \bibfield  {author} {\bibinfo {author} {\bibfnamefont {C.~W.}\ \bibnamefont
  {Bauer}}, \bibinfo {author} {\bibfnamefont {Z.}~\bibnamefont {Davoudi}},
  \bibinfo {author} {\bibfnamefont {A.~B.}\ \bibnamefont {Balantekin}},
  \bibinfo {author} {\bibfnamefont {T.}~\bibnamefont {Bhattacharya}}, \bibinfo
  {author} {\bibfnamefont {M.}~\bibnamefont {Carena}}, \bibinfo {author}
  {\bibfnamefont {W.~A.}\ \bibnamefont {de~Jong}}, \bibinfo {author}
  {\bibfnamefont {P.}~\bibnamefont {Draper}}, \bibinfo {author} {\bibfnamefont
  {A.}~\bibnamefont {El-Khadra}}, \bibinfo {author} {\bibfnamefont
  {N.}~\bibnamefont {Gemelke}}, \bibinfo {author} {\bibfnamefont
  {M.}~\bibnamefont {Hanada}}, \bibinfo {author} {\bibfnamefont
  {D.}~\bibnamefont {Kharzeev}}, \bibinfo {author} {\bibfnamefont
  {H.}~\bibnamefont {Lamm}}, \bibinfo {author} {\bibfnamefont {Y.-Y.}\
  \bibnamefont {Li}}, \bibinfo {author} {\bibfnamefont {J.}~\bibnamefont
  {Liu}}, \bibinfo {author} {\bibfnamefont {M.}~\bibnamefont {Lukin}}, \bibinfo
  {author} {\bibfnamefont {Y.}~\bibnamefont {Meurice}}, \bibinfo {author}
  {\bibfnamefont {C.}~\bibnamefont {Monroe}}, \bibinfo {author} {\bibfnamefont
  {B.}~\bibnamefont {Nachman}}, \bibinfo {author} {\bibfnamefont
  {G.}~\bibnamefont {Pagano}}, \bibinfo {author} {\bibfnamefont
  {J.}~\bibnamefont {Preskill}}, \bibinfo {author} {\bibfnamefont
  {E.}~\bibnamefont {Rinaldi}}, \bibinfo {author} {\bibfnamefont
  {A.}~\bibnamefont {Roggero}}, \bibinfo {author} {\bibfnamefont {D.~I.}\
  \bibnamefont {Santiago}}, \bibinfo {author} {\bibfnamefont {M.~J.}\
  \bibnamefont {Savage}}, \bibinfo {author} {\bibfnamefont {I.}~\bibnamefont
  {Siddiqi}}, \bibinfo {author} {\bibfnamefont {G.}~\bibnamefont {Siopsis}},
  \bibinfo {author} {\bibfnamefont {D.}~\bibnamefont {Van~Zanten}}, \bibinfo
  {author} {\bibfnamefont {N.}~\bibnamefont {Wiebe}}, \bibinfo {author}
  {\bibfnamefont {Y.}~\bibnamefont {Yamauchi}}, \bibinfo {author}
  {\bibfnamefont {K.}~\bibnamefont {Yeter-Aydeniz}},\ and\ \bibinfo {author}
  {\bibfnamefont {S.}~\bibnamefont {Zorzetti}},\ }\bibfield  {title} {\bibinfo
  {title} {Quantum simulation for high-energy physics},\ }\href
  {https://doi.org/10.1103/PRXQuantum.4.027001} {\bibfield  {journal} {\bibinfo
   {journal} {PRX Quantum}\ }\textbf {\bibinfo {volume} {4}},\ \bibinfo {pages}
  {027001} (\bibinfo {year} {2023})}\BibitemShut {NoStop}%
\bibitem [{\citenamefont {Di~Meglio}\ \emph {et~al.}(2024)\citenamefont
  {Di~Meglio}, \citenamefont {Jansen}, \citenamefont {Tavernelli},
  \citenamefont {Alexandrou}, \citenamefont {Arunachalam}, \citenamefont
  {Bauer}, \citenamefont {Borras}, \citenamefont {Carrazza}, \citenamefont
  {Crippa}, \citenamefont {Croft}, \citenamefont {de~Putter}, \citenamefont
  {Delgado}, \citenamefont {Dunjko}, \citenamefont {Egger}, \citenamefont
  {Fern\'andez-Combarro}, \citenamefont {Fuchs}, \citenamefont {Funcke},
  \citenamefont {Gonz\'alez-Cuadra}, \citenamefont {Grossi}, \citenamefont
  {Halimeh}, \citenamefont {Holmes}, \citenamefont {K\"uhn}, \citenamefont
  {Lacroix}, \citenamefont {Lewis}, \citenamefont {Lucchesi}, \citenamefont
  {Martinez}, \citenamefont {Meloni}, \citenamefont {Mezzacapo}, \citenamefont
  {Montangero}, \citenamefont {Nagano}, \citenamefont {Pascuzzi}, \citenamefont
  {Radescu}, \citenamefont {Ortega}, \citenamefont {Roggero}, \citenamefont
  {Schuhmacher}, \citenamefont {Seixas}, \citenamefont {Silvi}, \citenamefont
  {Spentzouris}, \citenamefont {Tacchino}, \citenamefont {Temme}, \citenamefont
  {Terashi}, \citenamefont {Tura}, \citenamefont {T\"uys\"uz}, \citenamefont
  {Vallecorsa}, \citenamefont {Wiese}, \citenamefont {Yoo},\ and\ \citenamefont
  {Zhang}}]{PRXQuantum.5.037001}%
  \BibitemOpen
  \bibfield  {author} {\bibinfo {author} {\bibfnamefont {A.}~\bibnamefont
  {Di~Meglio}}, \bibinfo {author} {\bibfnamefont {K.}~\bibnamefont {Jansen}},
  \bibinfo {author} {\bibfnamefont {I.}~\bibnamefont {Tavernelli}}, \bibinfo
  {author} {\bibfnamefont {C.}~\bibnamefont {Alexandrou}}, \bibinfo {author}
  {\bibfnamefont {S.}~\bibnamefont {Arunachalam}}, \bibinfo {author}
  {\bibfnamefont {C.~W.}\ \bibnamefont {Bauer}}, \bibinfo {author}
  {\bibfnamefont {K.}~\bibnamefont {Borras}}, \bibinfo {author} {\bibfnamefont
  {S.}~\bibnamefont {Carrazza}}, \bibinfo {author} {\bibfnamefont
  {A.}~\bibnamefont {Crippa}}, \bibinfo {author} {\bibfnamefont
  {V.}~\bibnamefont {Croft}}, \bibinfo {author} {\bibfnamefont
  {R.}~\bibnamefont {de~Putter}}, \bibinfo {author} {\bibfnamefont
  {A.}~\bibnamefont {Delgado}}, \bibinfo {author} {\bibfnamefont
  {V.}~\bibnamefont {Dunjko}}, \bibinfo {author} {\bibfnamefont {D.~J.}\
  \bibnamefont {Egger}}, \bibinfo {author} {\bibfnamefont {E.}~\bibnamefont
  {Fern\'andez-Combarro}}, \bibinfo {author} {\bibfnamefont {E.}~\bibnamefont
  {Fuchs}}, \bibinfo {author} {\bibfnamefont {L.}~\bibnamefont {Funcke}},
  \bibinfo {author} {\bibfnamefont {D.}~\bibnamefont {Gonz\'alez-Cuadra}},
  \bibinfo {author} {\bibfnamefont {M.}~\bibnamefont {Grossi}}, \bibinfo
  {author} {\bibfnamefont {J.~C.}\ \bibnamefont {Halimeh}}, \bibinfo {author}
  {\bibfnamefont {Z.}~\bibnamefont {Holmes}}, \bibinfo {author} {\bibfnamefont
  {S.}~\bibnamefont {K\"uhn}}, \bibinfo {author} {\bibfnamefont
  {D.}~\bibnamefont {Lacroix}}, \bibinfo {author} {\bibfnamefont
  {R.}~\bibnamefont {Lewis}}, \bibinfo {author} {\bibfnamefont
  {D.}~\bibnamefont {Lucchesi}}, \bibinfo {author} {\bibfnamefont {M.~L.}\
  \bibnamefont {Martinez}}, \bibinfo {author} {\bibfnamefont {F.}~\bibnamefont
  {Meloni}}, \bibinfo {author} {\bibfnamefont {A.}~\bibnamefont {Mezzacapo}},
  \bibinfo {author} {\bibfnamefont {S.}~\bibnamefont {Montangero}}, \bibinfo
  {author} {\bibfnamefont {L.}~\bibnamefont {Nagano}}, \bibinfo {author}
  {\bibfnamefont {V.~R.}\ \bibnamefont {Pascuzzi}}, \bibinfo {author}
  {\bibfnamefont {V.}~\bibnamefont {Radescu}}, \bibinfo {author} {\bibfnamefont
  {E.~R.}\ \bibnamefont {Ortega}}, \bibinfo {author} {\bibfnamefont
  {A.}~\bibnamefont {Roggero}}, \bibinfo {author} {\bibfnamefont
  {J.}~\bibnamefont {Schuhmacher}}, \bibinfo {author} {\bibfnamefont
  {J.}~\bibnamefont {Seixas}}, \bibinfo {author} {\bibfnamefont
  {P.}~\bibnamefont {Silvi}}, \bibinfo {author} {\bibfnamefont
  {P.}~\bibnamefont {Spentzouris}}, \bibinfo {author} {\bibfnamefont
  {F.}~\bibnamefont {Tacchino}}, \bibinfo {author} {\bibfnamefont
  {K.}~\bibnamefont {Temme}}, \bibinfo {author} {\bibfnamefont
  {K.}~\bibnamefont {Terashi}}, \bibinfo {author} {\bibfnamefont
  {J.}~\bibnamefont {Tura}}, \bibinfo {author} {\bibfnamefont {C.}~\bibnamefont
  {T\"uys\"uz}}, \bibinfo {author} {\bibfnamefont {S.}~\bibnamefont
  {Vallecorsa}}, \bibinfo {author} {\bibfnamefont {U.-J.}\ \bibnamefont
  {Wiese}}, \bibinfo {author} {\bibfnamefont {S.}~\bibnamefont {Yoo}},\ and\
  \bibinfo {author} {\bibfnamefont {J.}~\bibnamefont {Zhang}},\ }\bibfield
  {title} {\bibinfo {title} {Quantum computing for high-energy physics: State
  of the art and challenges},\ }\href
  {https://doi.org/10.1103/PRXQuantum.5.037001} {\bibfield  {journal} {\bibinfo
   {journal} {PRX Quantum}\ }\textbf {\bibinfo {volume} {5}},\ \bibinfo {pages}
  {037001} (\bibinfo {year} {2024})}\BibitemShut {NoStop}%
\bibitem [{\citenamefont {Halimeh}\ \emph {et~al.}(2022)\citenamefont
  {Halimeh}, \citenamefont {McCulloch}, \citenamefont {Yang},\ and\
  \citenamefont {Hauke}}]{PRXQuantum.3.040316}%
  \BibitemOpen
  \bibfield  {author} {\bibinfo {author} {\bibfnamefont {J.~C.}\ \bibnamefont
  {Halimeh}}, \bibinfo {author} {\bibfnamefont {I.~P.}\ \bibnamefont
  {McCulloch}}, \bibinfo {author} {\bibfnamefont {B.}~\bibnamefont {Yang}},\
  and\ \bibinfo {author} {\bibfnamefont {P.}~\bibnamefont {Hauke}},\ }\bibfield
   {title} {\bibinfo {title} {Tuning the topological
  $\ensuremath{\theta}$-angle in cold-atom quantum simulators of gauge
  theories},\ }\href {https://doi.org/10.1103/PRXQuantum.3.040316} {\bibfield
  {journal} {\bibinfo  {journal} {PRX Quantum}\ }\textbf {\bibinfo {volume}
  {3}},\ \bibinfo {pages} {040316} (\bibinfo {year} {2022})}\BibitemShut
  {NoStop}%
\bibitem [{\citenamefont {Zhang}\ \emph {et~al.}(2025)\citenamefont {Zhang},
  \citenamefont {Liu}, \citenamefont {Cheng}, \citenamefont {He}, \citenamefont
  {Wang}, \citenamefont {Wang}, \citenamefont {Zhu}, \citenamefont {Su},
  \citenamefont {Zhou}, \citenamefont {Zheng}, \citenamefont {Sun},
  \citenamefont {Yang}, \citenamefont {Hauke}, \citenamefont {Zheng},
  \citenamefont {Halimeh}, \citenamefont {Yuan},\ and\ \citenamefont
  {Pan}}]{Zhang2025}%
  \BibitemOpen
  \bibfield  {author} {\bibinfo {author} {\bibfnamefont {W.-Y.}\ \bibnamefont
  {Zhang}}, \bibinfo {author} {\bibfnamefont {Y.}~\bibnamefont {Liu}}, \bibinfo
  {author} {\bibfnamefont {Y.}~\bibnamefont {Cheng}}, \bibinfo {author}
  {\bibfnamefont {M.-G.}\ \bibnamefont {He}}, \bibinfo {author} {\bibfnamefont
  {H.-Y.}\ \bibnamefont {Wang}}, \bibinfo {author} {\bibfnamefont {T.-Y.}\
  \bibnamefont {Wang}}, \bibinfo {author} {\bibfnamefont {Z.-H.}\ \bibnamefont
  {Zhu}}, \bibinfo {author} {\bibfnamefont {G.-X.}\ \bibnamefont {Su}},
  \bibinfo {author} {\bibfnamefont {Z.-Y.}\ \bibnamefont {Zhou}}, \bibinfo
  {author} {\bibfnamefont {Y.-G.}\ \bibnamefont {Zheng}}, \bibinfo {author}
  {\bibfnamefont {H.}~\bibnamefont {Sun}}, \bibinfo {author} {\bibfnamefont
  {B.}~\bibnamefont {Yang}}, \bibinfo {author} {\bibfnamefont {P.}~\bibnamefont
  {Hauke}}, \bibinfo {author} {\bibfnamefont {W.}~\bibnamefont {Zheng}},
  \bibinfo {author} {\bibfnamefont {J.~C.}\ \bibnamefont {Halimeh}}, \bibinfo
  {author} {\bibfnamefont {Z.-S.}\ \bibnamefont {Yuan}},\ and\ \bibinfo
  {author} {\bibfnamefont {J.-W.}\ \bibnamefont {Pan}},\ }\bibfield  {title}
  {\bibinfo {title} {Observation of microscopic confinement dynamics by a
  tunable topological $\theta$-angle},\ }\href
  {https://doi.org/10.1038/s41567-024-02702-x} {\bibfield  {journal} {\bibinfo
  {journal} {Nature Physics}\ }\textbf {\bibinfo {volume} {21}},\ \bibinfo
  {pages} {155} (\bibinfo {year} {2025})}\BibitemShut {NoStop}%
\bibitem [{\citenamefont {Magann}\ \emph {et~al.}(2022)\citenamefont {Magann},
  \citenamefont {Rudinger}, \citenamefont {Grace},\ and\ \citenamefont
  {Sarovar}}]{PhysRevLett.129.250502}%
  \BibitemOpen
  \bibfield  {author} {\bibinfo {author} {\bibfnamefont {A.~B.}\ \bibnamefont
  {Magann}}, \bibinfo {author} {\bibfnamefont {K.~M.}\ \bibnamefont
  {Rudinger}}, \bibinfo {author} {\bibfnamefont {M.~D.}\ \bibnamefont
  {Grace}},\ and\ \bibinfo {author} {\bibfnamefont {M.}~\bibnamefont
  {Sarovar}},\ }\bibfield  {title} {\bibinfo {title} {Feedback-based quantum
  optimization},\ }\href {https://doi.org/10.1103/PhysRevLett.129.250502}
  {\bibfield  {journal} {\bibinfo  {journal} {Phys. Rev. Lett.}\ }\textbf
  {\bibinfo {volume} {129}},\ \bibinfo {pages} {250502} (\bibinfo {year}
  {2022})}\BibitemShut {NoStop}%
\bibitem [{\citenamefont {Nguyen Van~Long}\ \emph {et~al.}(2025)\citenamefont
  {Nguyen Van~Long}, \citenamefont {Nguyen~Tran},\ and\ \citenamefont
  {Ho}}]{nguyen2025imaginary}%
  \BibitemOpen
  \bibfield  {author} {\bibinfo {author} {\bibfnamefont {T.}~\bibnamefont
  {Nguyen Van~Long}}, \bibinfo {author} {\bibfnamefont {L.}~\bibnamefont
  {Nguyen~Tran}},\ and\ \bibinfo {author} {\bibfnamefont {L.~B.}\ \bibnamefont
  {Ho}},\ }\bibfield  {title} {\bibinfo {title} {Imaginary-time-enhanced
  feedback-based quantum algorithms for universal ground-state preparation},\
  }\href@noop {} {\bibfield  {journal} {\bibinfo  {journal} {arXiv e-prints}\
  ,\ \bibinfo {pages} {arXiv}} (\bibinfo {year} {2025})}\BibitemShut {NoStop}%
\bibitem [{\citenamefont {Zia}\ \emph {et~al.}(2009)\citenamefont {Zia},
  \citenamefont {Redish},\ and\ \citenamefont {McKay}}]{10.1119/1.3119512}%
  \BibitemOpen
  \bibfield  {author} {\bibinfo {author} {\bibfnamefont {R.~K.~P.}\
  \bibnamefont {Zia}}, \bibinfo {author} {\bibfnamefont {E.~F.}\ \bibnamefont
  {Redish}},\ and\ \bibinfo {author} {\bibfnamefont {S.~R.}\ \bibnamefont
  {McKay}},\ }\bibfield  {title} {\bibinfo {title} {Making sense of the
  legendre transform},\ }\href {https://doi.org/10.1119/1.3119512} {\bibfield
  {journal} {\bibinfo  {journal} {American Journal of Physics}\ }\textbf
  {\bibinfo {volume} {77}},\ \bibinfo {pages} {614} (\bibinfo {year}
  {2009})}\BibitemShut {NoStop}%
\bibitem [{\citenamefont {Aoki}\ and\ \citenamefont
  {Ichihara}(1995)}]{PhysRevD.52.6435}%
  \BibitemOpen
  \bibfield  {author} {\bibinfo {author} {\bibfnamefont {K.}~\bibnamefont
  {Aoki}}\ and\ \bibinfo {author} {\bibfnamefont {T.}~\bibnamefont
  {Ichihara}},\ }\bibfield  {title} {\bibinfo {title} {(1+1)-dimensional qcd
  with fundamental bosons and fermions},\ }\href
  {https://doi.org/10.1103/PhysRevD.52.6435} {\bibfield  {journal} {\bibinfo
  {journal} {Phys. Rev. D}\ }\textbf {\bibinfo {volume} {52}},\ \bibinfo
  {pages} {6435} (\bibinfo {year} {1995})}\BibitemShut {NoStop}%
\bibitem [{\citenamefont {Schwinger}(1962)}]{PhysRev.128.2425}%
  \BibitemOpen
  \bibfield  {author} {\bibinfo {author} {\bibfnamefont {J.}~\bibnamefont
  {Schwinger}},\ }\bibfield  {title} {\bibinfo {title} {Gauge invariance and
  mass. ii},\ }\href {https://doi.org/10.1103/PhysRev.128.2425} {\bibfield
  {journal} {\bibinfo  {journal} {Phys. Rev.}\ }\textbf {\bibinfo {volume}
  {128}},\ \bibinfo {pages} {2425} (\bibinfo {year} {1962})}\BibitemShut
  {NoStop}%
\bibitem [{\citenamefont {Luo}(2007)}]{Luo2007}%
  \BibitemOpen
  \bibfield  {author} {\bibinfo {author} {\bibfnamefont {X.}~\bibnamefont
  {Luo}},\ }\bibfield  {title} {\bibinfo {title} {Spontaneous chiral-symmetry
  breaking of lattice qcd with massless dynamical quarks},\ }\href
  {https://doi.org/10.1007/s11433-007-2015-5} {\bibfield  {journal} {\bibinfo
  {journal} {Science in China Series G: Physics, Mechanics and Astronomy}\
  }\textbf {\bibinfo {volume} {50}},\ \bibinfo {pages} {6} (\bibinfo {year}
  {2007})}\BibitemShut {NoStop}%
\bibitem [{\citenamefont {Ballon-Bayona}\ \emph {et~al.}(2021)\citenamefont
  {Ballon-Bayona}, \citenamefont {Mamani},\ and\ \citenamefont
  {Rodrigues}}]{PhysRevD.104.126029}%
  \BibitemOpen
  \bibfield  {author} {\bibinfo {author} {\bibfnamefont {A.}~\bibnamefont
  {Ballon-Bayona}}, \bibinfo {author} {\bibfnamefont {L.~A.~H.}\ \bibnamefont
  {Mamani}},\ and\ \bibinfo {author} {\bibfnamefont {D.~M.}\ \bibnamefont
  {Rodrigues}},\ }\bibfield  {title} {\bibinfo {title} {Spontaneous chiral
  symmetry breaking in holographic soft wall models},\ }\href
  {https://doi.org/10.1103/PhysRevD.104.126029} {\bibfield  {journal} {\bibinfo
   {journal} {Phys. Rev. D}\ }\textbf {\bibinfo {volume} {104}},\ \bibinfo
  {pages} {126029} (\bibinfo {year} {2021})}\BibitemShut {NoStop}%
\bibitem [{\citenamefont {Funcke}\ \emph {et~al.}(2020)\citenamefont {Funcke},
  \citenamefont {Jansen},\ and\ \citenamefont {K\"uhn}}]{PhysRevD.101.054507}%
  \BibitemOpen
  \bibfield  {author} {\bibinfo {author} {\bibfnamefont {L.}~\bibnamefont
  {Funcke}}, \bibinfo {author} {\bibfnamefont {K.}~\bibnamefont {Jansen}},\
  and\ \bibinfo {author} {\bibfnamefont {S.}~\bibnamefont {K\"uhn}},\
  }\bibfield  {title} {\bibinfo {title} {Topological vacuum structure of the
  schwinger model with matrix product states},\ }\href
  {https://doi.org/10.1103/PhysRevD.101.054507} {\bibfield  {journal} {\bibinfo
   {journal} {Phys. Rev. D}\ }\textbf {\bibinfo {volume} {101}},\ \bibinfo
  {pages} {054507} (\bibinfo {year} {2020})}\BibitemShut {NoStop}%
\bibitem [{\citenamefont {Smilga}(1992)}]{PhysRevD.46.5598}%
  \BibitemOpen
  \bibfield  {author} {\bibinfo {author} {\bibfnamefont {A.~V.}\ \bibnamefont
  {Smilga}},\ }\bibfield  {title} {\bibinfo {title} {Vacuum fields in the
  schwinger model},\ }\href {https://doi.org/10.1103/PhysRevD.46.5598}
  {\bibfield  {journal} {\bibinfo  {journal} {Phys. Rev. D}\ }\textbf {\bibinfo
  {volume} {46}},\ \bibinfo {pages} {5598} (\bibinfo {year}
  {1992})}\BibitemShut {NoStop}%
\bibitem [{\citenamefont {Calliari}\ \emph {et~al.}(2025)\citenamefont
  {Calliari}, \citenamefont {Di~Liberto}, \citenamefont {Pichler},\ and\
  \citenamefont {Zache}}]{nt76-ttmj}%
  \BibitemOpen
  \bibfield  {author} {\bibinfo {author} {\bibfnamefont {G.}~\bibnamefont
  {Calliari}}, \bibinfo {author} {\bibfnamefont {M.}~\bibnamefont
  {Di~Liberto}}, \bibinfo {author} {\bibfnamefont {H.}~\bibnamefont
  {Pichler}},\ and\ \bibinfo {author} {\bibfnamefont {T.~V.}\ \bibnamefont
  {Zache}},\ }\bibfield  {title} {\bibinfo {title} {Quantum simulating
  continuum field theories with large-spin lattice models},\ }\href
  {https://doi.org/10.1103/nt76-ttmj} {\bibfield  {journal} {\bibinfo
  {journal} {PRX Quantum}\ }\textbf {\bibinfo {volume} {6}},\ \bibinfo {pages}
  {030304} (\bibinfo {year} {2025})}\BibitemShut {NoStop}%
\bibitem [{\citenamefont {Altman}\ \emph {et~al.}(2021)\citenamefont {Altman},
  \citenamefont {Brown}, \citenamefont {Carleo}, \citenamefont {Carr},
  \citenamefont {Demler}, \citenamefont {Chin}, \citenamefont {DeMarco},
  \citenamefont {Economou}, \citenamefont {Eriksson}, \citenamefont {Fu},
  \citenamefont {Greiner}, \citenamefont {Hazzard}, \citenamefont {Hulet},
  \citenamefont {Koll\'ar}, \citenamefont {Lev}, \citenamefont {Lukin},
  \citenamefont {Ma}, \citenamefont {Mi}, \citenamefont {Misra}, \citenamefont
  {Monroe}, \citenamefont {Murch}, \citenamefont {Nazario}, \citenamefont {Ni},
  \citenamefont {Potter}, \citenamefont {Roushan}, \citenamefont {Saffman},
  \citenamefont {Schleier-Smith}, \citenamefont {Siddiqi}, \citenamefont
  {Simmonds}, \citenamefont {Singh}, \citenamefont {Spielman}, \citenamefont
  {Temme}, \citenamefont {Weiss}, \citenamefont {Vu\ifmmode \check{c}\else
  \v{c}\fi{}kovi\ifmmode~\acute{c}\else \'{c}\fi{}}, \citenamefont
  {Vuleti\ifmmode~\acute{c}\else \'{c}\fi{}}, \citenamefont {Ye},\ and\
  \citenamefont {Zwierlein}}]{PRXQuantum.2.017003}%
  \BibitemOpen
  \bibfield  {author} {\bibinfo {author} {\bibfnamefont {E.}~\bibnamefont
  {Altman}}, \bibinfo {author} {\bibfnamefont {K.~R.}\ \bibnamefont {Brown}},
  \bibinfo {author} {\bibfnamefont {G.}~\bibnamefont {Carleo}}, \bibinfo
  {author} {\bibfnamefont {L.~D.}\ \bibnamefont {Carr}}, \bibinfo {author}
  {\bibfnamefont {E.}~\bibnamefont {Demler}}, \bibinfo {author} {\bibfnamefont
  {C.}~\bibnamefont {Chin}}, \bibinfo {author} {\bibfnamefont {B.}~\bibnamefont
  {DeMarco}}, \bibinfo {author} {\bibfnamefont {S.~E.}\ \bibnamefont
  {Economou}}, \bibinfo {author} {\bibfnamefont {M.~A.}\ \bibnamefont
  {Eriksson}}, \bibinfo {author} {\bibfnamefont {K.-M.~C.}\ \bibnamefont {Fu}},
  \bibinfo {author} {\bibfnamefont {M.}~\bibnamefont {Greiner}}, \bibinfo
  {author} {\bibfnamefont {K.~R.}\ \bibnamefont {Hazzard}}, \bibinfo {author}
  {\bibfnamefont {R.~G.}\ \bibnamefont {Hulet}}, \bibinfo {author}
  {\bibfnamefont {A.~J.}\ \bibnamefont {Koll\'ar}}, \bibinfo {author}
  {\bibfnamefont {B.~L.}\ \bibnamefont {Lev}}, \bibinfo {author} {\bibfnamefont
  {M.~D.}\ \bibnamefont {Lukin}}, \bibinfo {author} {\bibfnamefont
  {R.}~\bibnamefont {Ma}}, \bibinfo {author} {\bibfnamefont {X.}~\bibnamefont
  {Mi}}, \bibinfo {author} {\bibfnamefont {S.}~\bibnamefont {Misra}}, \bibinfo
  {author} {\bibfnamefont {C.}~\bibnamefont {Monroe}}, \bibinfo {author}
  {\bibfnamefont {K.}~\bibnamefont {Murch}}, \bibinfo {author} {\bibfnamefont
  {Z.}~\bibnamefont {Nazario}}, \bibinfo {author} {\bibfnamefont {K.-K.}\
  \bibnamefont {Ni}}, \bibinfo {author} {\bibfnamefont {A.~C.}\ \bibnamefont
  {Potter}}, \bibinfo {author} {\bibfnamefont {P.}~\bibnamefont {Roushan}},
  \bibinfo {author} {\bibfnamefont {M.}~\bibnamefont {Saffman}}, \bibinfo
  {author} {\bibfnamefont {M.}~\bibnamefont {Schleier-Smith}}, \bibinfo
  {author} {\bibfnamefont {I.}~\bibnamefont {Siddiqi}}, \bibinfo {author}
  {\bibfnamefont {R.}~\bibnamefont {Simmonds}}, \bibinfo {author}
  {\bibfnamefont {M.}~\bibnamefont {Singh}}, \bibinfo {author} {\bibfnamefont
  {I.}~\bibnamefont {Spielman}}, \bibinfo {author} {\bibfnamefont
  {K.}~\bibnamefont {Temme}}, \bibinfo {author} {\bibfnamefont {D.~S.}\
  \bibnamefont {Weiss}}, \bibinfo {author} {\bibfnamefont {J.}~\bibnamefont
  {Vu\ifmmode \check{c}\else \v{c}\fi{}kovi\ifmmode~\acute{c}\else
  \'{c}\fi{}}}, \bibinfo {author} {\bibfnamefont {V.}~\bibnamefont
  {Vuleti\ifmmode~\acute{c}\else \'{c}\fi{}}}, \bibinfo {author} {\bibfnamefont
  {J.}~\bibnamefont {Ye}},\ and\ \bibinfo {author} {\bibfnamefont
  {M.}~\bibnamefont {Zwierlein}},\ }\bibfield  {title} {\bibinfo {title}
  {Quantum simulators: Architectures and opportunities},\ }\href
  {https://doi.org/10.1103/PRXQuantum.2.017003} {\bibfield  {journal} {\bibinfo
   {journal} {PRX Quantum}\ }\textbf {\bibinfo {volume} {2}},\ \bibinfo {pages}
  {017003} (\bibinfo {year} {2021})}\BibitemShut {NoStop}%
\bibitem [{\citenamefont {Kogut}\ and\ \citenamefont
  {Susskind}(1975)}]{PhysRevD.11.395}%
  \BibitemOpen
  \bibfield  {author} {\bibinfo {author} {\bibfnamefont {J.}~\bibnamefont
  {Kogut}}\ and\ \bibinfo {author} {\bibfnamefont {L.}~\bibnamefont
  {Susskind}},\ }\bibfield  {title} {\bibinfo {title} {Hamiltonian formulation
  of wilson's lattice gauge theories},\ }\href
  {https://doi.org/10.1103/PhysRevD.11.395} {\bibfield  {journal} {\bibinfo
  {journal} {Phys. Rev. D}\ }\textbf {\bibinfo {volume} {11}},\ \bibinfo
  {pages} {395} (\bibinfo {year} {1975})}\BibitemShut {NoStop}%
\bibitem [{\citenamefont {Zohar}\ and\ \citenamefont
  {Burrello}(2015)}]{PhysRevD.91.054506}%
  \BibitemOpen
  \bibfield  {author} {\bibinfo {author} {\bibfnamefont {E.}~\bibnamefont
  {Zohar}}\ and\ \bibinfo {author} {\bibfnamefont {M.}~\bibnamefont
  {Burrello}},\ }\bibfield  {title} {\bibinfo {title} {Formulation of lattice
  gauge theories for quantum simulations},\ }\href
  {https://doi.org/10.1103/PhysRevD.91.054506} {\bibfield  {journal} {\bibinfo
  {journal} {Phys. Rev. D}\ }\textbf {\bibinfo {volume} {91}},\ \bibinfo
  {pages} {054506} (\bibinfo {year} {2015})}\BibitemShut {NoStop}%
\bibitem [{\citenamefont {Jordan}\ and\ \citenamefont
  {Wigner}(1928)}]{Jordan1928}%
  \BibitemOpen
  \bibfield  {author} {\bibinfo {author} {\bibfnamefont {P.}~\bibnamefont
  {Jordan}}\ and\ \bibinfo {author} {\bibfnamefont {E.}~\bibnamefont
  {Wigner}},\ }\bibfield  {title} {\bibinfo {title} {{\"U}ber das paulische
  {\"a}quivalenzverbot},\ }\href {https://doi.org/10.1007/BF01331938}
  {\bibfield  {journal} {\bibinfo  {journal} {Zeitschrift f{\"u}r Physik}\
  }\textbf {\bibinfo {volume} {47}},\ \bibinfo {pages} {631} (\bibinfo {year}
  {1928})}\BibitemShut {NoStop}%
\bibitem [{\citenamefont {Ho}(2026)}]{strongcp}%
  \BibitemOpen
  \bibfield  {author} {\bibinfo {author} {\bibfnamefont {L.~B.}\ \bibnamefont
  {Ho}},\ }\href@noop {} {\bibinfo {title} {Supporting data for quantum
  simulation of strong charge-parity violation and the peccei-quinn
  mechanism}},\ \bibinfo {howpublished}
  {\url{https://github.com/echkon/strongCP}} (\bibinfo {year}
  {2026})\BibitemShut {NoStop}%
\end{thebibliography}%
\end{document}